\documentclass[12pt,a4paper]{article}
\usepackage{graphicx}
\usepackage{dcolumn}
\usepackage{bm}
\usepackage{hyperref}
\usepackage{braket}
\usepackage{authblk}
\usepackage{float}
\usepackage{mathptmx}  
\usepackage{epstopdf}
\usepackage{fancyhdr}
\begin{document}
\title{Position Measurement-Induced Collapse: \\ A Unified Quantum  Description of Fraunhofer and Fresnel Diffractions}
\author[1]{ Moncy V. John\thanks{moncyjohn@yahoo.co.uk}}
\author[1,2]{Kiran Mathew \thanks{kiran007x@yahoo.co.in}}
\affil[1]{ Department of Physics, St. Thomas College, Kozhencherri - 689641, Kerala, India}
 \affil[2]{ Department of Physics, Baselius College, Kottayam-686001, Kerala, India}
\date{\today}

\maketitle
\begin{abstract}
Position measurement-induced collapse  states are shown to provide a unified quantum description of diffraction of  particles passing through a single slit. These states, which we here call `quantum location states', are  represented by the conventional rectangular wave function at the initial moment of  position measurement. We expand this state in terms of the position eigenstates, which in turn can be represented as a linear combination of energy eigenfunctions of the problem, using the closure property. The time-evolution of the location  states in the case of free particles is shown to have position probability density patterns closely resembling diffraction patterns in  the  Fresnel region for small times and the same in Fraunhofer region for large times.  Using the quantum trajectory representations in the de Broglie-Bohm, modified de Broglie-Bohm and Floyd-Faraggi-Matone formalisms, we show that Fresnel and Fraunhofer diffractions can be   described using a single expression. We  also discuss how to obtain the probability density of location states for the  case of particles moving in a general potential, detected at some arbitrary point. In the case of  the harmonic oscillator potential, we find that they   have oscillatory properties  similar to that of coherent states.

\end{abstract}

\section{Introduction}
An early attempt  to describe   the Fraunhoffer diffraction in optics using quantum theory  was made by  Epstein and Ehrenfest  \cite{epstein},  by combining the concept of a light quantum with Bohr's correspondence principle.   However, a consistent quantum treatment of  the  diffraction of  particles passing  through a single slit is not yet at hand and it is  pointed out (see, for example, \cite{sudarshan}) that the usual textbook discussions of such phenomena are fraught with ambiguities and omissions. This observation is based mainly on the fact  that  such  treatises   directly use the  classical  wave optics itself, instead of making a quantum mechanical approach, to explain diffraction, interference etc.  This fundamental issue is addressed in some recent works \cite{marcella,rothman,fabbro},  and the authors have proceeded further to explore the possibility  that the diffraction through a single slit can be regarded as a prototype experiment to all position measurements. That is, they considered the slit in the diaphragm as a device for measuring the position of the incident particle,  hoping that such an analysis  shall be helpful for  a general understanding of the quantum theory of position measurement. It is also noted that in the literature, the only calculations which treat diffraction through  single slit as `position measurement' are  those cited above. 

 Among these  works, the earliest one  due to Marcella \cite{marcella}  proposes an expression to calculate the intensity of the diffracted beam   using a reduced  state ket of the particle. The collapse of state due to the process of measurement  leads to this  wave function, which is equivalently   the   projection of an initial state ket into the final one. The Fourier transform of the resulting function, which is simply the momentum space wave function in the same case, is squared to give an expression that is comparable with the standard result in Fraunhofer diffraction.   Rothman and Boughn \cite{rothman} have shown that this procedure in \cite{marcella} is not compatible with standard quantum mechanics. Recently,  Fabbro \cite{fabbro}  has come up with an extension of the formalism in \cite{marcella}, in an attempt to  rectify its drawbacks. Here it is inferred that these drawbacks  come from the way in which the wave function reduction has been applied in it. The model in \cite{fabbro} claims to predict the intensity of diffracted wave at large angles, focusing on the case of diffraction at infinity (Fraunhofer diffraction) only.  
 
 In the present work, we intend to study this issue  and to suggest  an alternative approach to that in \cite{marcella}. In the first part, the general case  of performing a measurement of position on a particle in one dimension is taken up. Initially, we consider a free particle and later a particle under an arbitrary potential. When a position measurement  takes place at a point,  it is assumed to lead to a collapse of the original wave function.  Following the conventional approach in orthogonal measurements \cite{kip,wise}, we assume the collapse to be in such a way that the  resulting  function  is a nonzero constant  inside the region of a  one-dimensional box centred about the point and zero outside it. Though  this collapsed state is not  a position eigenstate, we can consider it as a superposition of position eigenstates whose  eigenvalues  lie inside the box. Here  use is made of the fact that the position eigenstates have the general form of the  Dirac delta function.  The crucial steps to follow  are    the expansion of the above collapsed wave function by substituting the delta function with  the left hand side of the closure relation for energy eigenstates in this case, and the  introduction of the unitary time evolution of the collapsed state.  The resulting states fall into a  category of its own, which we here call `position measurement-induced collapse state' or `location state' for short. After developing such a method for the position measurement of a free  particle in one-dimension and later for a particle in a harmonic oscillator potential, we apply it to the phenomenon of diffraction of  particles  passing through  a single slit. Our numerical calculations demonstrate that the location states can play  an important  role  in explaining phenomena such as quantum diffraction and interference.

Another important element in the present work is the use of quantum trajectories for the description of diffraction. We agree with the  observation made by Rothman and Boughn that Ref. \cite{marcella} adopts a  hidden variable point of view, though it is in a very crude form. On the other hand, in this paper, we  use some standard nonlocal hidden variable theories for the quantum description of  diffraction. Historically,  hidden  variables  were viewed by quantum physicists   as simply part of  attempts to make  quantum mechanics a causal and local theory. More specifically, according to earlier conventional viewpoint, hidden variable theories cannot  violate the well-known Bell's inequalities \cite{bell}. However, Bell himself has  identified and publicised that one cannot rule out the existence of  nonlocal  hidden variable theories, such as the de Broglie-Bohm (dBB) theory \cite{dBB1,dBB2}. Recently,  modified de Broglie-Bohm (MdBB) \cite{MdBB,yang,tannor} and Floyd-Faraggi-Matone (FFM) \cite{floyd,matone} quantum trajectories, which are also nonlocal hidden variable theories, are widely discussed. These trajectory formalisms  give the same predictions as those  in standard quantum theory (except that of the trajectories, which are not directly measurable). Hence generally they   agree with the latter on all predictions of experimental results.  In this paper, we have  shown that quantum trajectories such as those in the dBB, MdBB and FFM formalisms can be of valuable help in explaining diffraction.  By this approach,  a unified description of the phenomenon can be obtained, applicable both in the Fresnel and Fraunhofer regions. 
 
The paper is organized as follows.  We start from the current studies \cite{marcella,rothman,fabbro} on the above problem of quantum diffraction in Sec. \ref{sec:review},  giving a careful account of the procedure  in \cite{marcella}. In Sec. \ref{sec:free}, the  perspective on the measurement of position adopted in this paper and the location state wave function  for a free particle in one dimension are presented. The corresponding problem when the particle is moving in an arbitrary potential   is discussed in Sec. \ref{sec:general}. How to  describe diffraction of particles with the help of both location states and  quantum trajectories is explained in Sec. \ref{sec:diffraction}. The last section comprises a discussion of our results.
 
\section{Diffraction as a measurement process} \label{sec:review}

\begin{figure}[h]
\centering 
\includegraphics[width=0.85\textwidth]{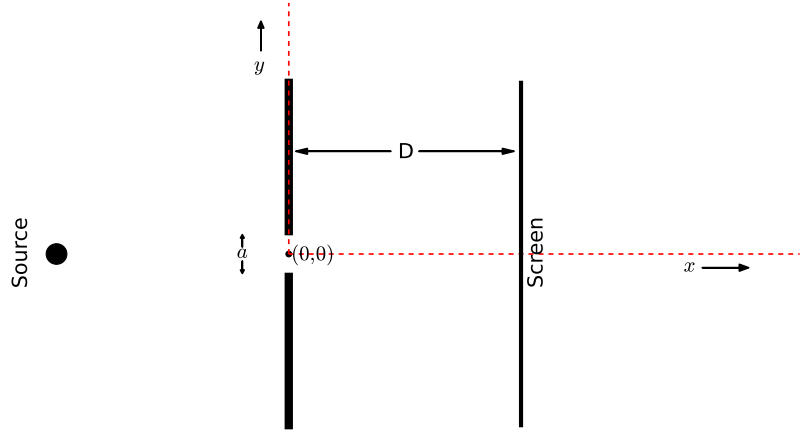}
\caption{Experimental set up}
\label{fig:apparatus}        
\end{figure}

 Here we briefly review the work in \cite{marcella} and its modifications suggested in \cite{fabbro}. A single-slit diffraction experiment with an apparatus  as shown in Fig. \ref{fig:apparatus} is considered.  On a diaphragm kept in the plane  $x=0$, a slit of width $\Delta y =a$  is made with center at $y=0$. Let the slit be  of infinite depth along the $z$-axis.  It is assumed that in the product wave function  of the   incident free particle, only  the part   along the $y$-axis is  affected by  diffraction. This part of   the initial state   gets reduced to  $\ket{\psi_y^{y=0,\Delta y=a} }$ at the moment of collapse. We denote this collapsed ket as $\ket{\psi_y^{0,a}}$ for short. The corresponding  quantum wave function of the particle in the position space is $\psi_y^{0,a} (y) \equiv \braket{y|\psi_y^{0,a} }$. In the momentum space, this state is described    by $ \braket{p_y|\psi_y^{0,a} }$, which may be denoted as    $\phi_y^{0,a}(p_y)$. The probability  that the particle is scattered with its $y$-momentum  between $p_y$ and $p_y + dp_y$ is then given by

\begin{equation}
P(p_y)dp_y=|\phi_y^{0,a}(p_y) |^2 dp_y. \label{eq:probpy}
\end{equation}  
Here, the wave function in the momentum space is related to the corresponding wave function in position space by

\begin{eqnarray}
\phi_y^{0,a}(p_y) \equiv  \braket{p_y|\psi_y^{0,a} } &= & \int dy \braket{p_y|y}\braket{y|\psi_y^{0,a} }  \nonumber \\
&=& \frac{1}{\sqrt{2 \pi}} \int dy \exp \left({\frac{-ip_yy}{\hbar}}\right) \psi_y^{0,a} (y) . \label{eq:mom_spc_wvfn}
\end{eqnarray}
 In the above,

\begin{equation}
\braket{p_y|y} =  \frac{1}{\sqrt{2 \pi}}  \exp\left({\frac{-ip_yy}{\hbar}}\right) ,
\end{equation}
 is the position eigenfunction in the  momentum space. Marcella assumes the  collapsed  wave function in position space, in a position measurement such as the above, to be of the normalised form

\begin{eqnarray}
\psi_y^{0,a}  (y)  =
\left\{
\begin{array}{lr}
\frac{1}{\sqrt{a}}  &\qquad -a/2\leq y\leq a/2\\
0& \qquad |y|>a/2
\end{array}
\right. ,
\label{eq:slit_wavefn1}
\end{eqnarray}
 at the instant of  measurement (passage through the slit).   The  momentum space wave function corresponding to this   may be evaluated using Eq. (\ref{eq:mom_spc_wvfn}) as 

\begin{equation}
 \phi_y^{0,a}(p_y)= \frac{2\hbar}{p_y\sqrt{2\pi a}}\sin\left( \frac{ap_y}{2\hbar}\right) , \label{eq:sync} 
\end{equation}
 which is the Fourier transform of (\ref{eq:slit_wavefn1}). The  probability density for this particle to have momentum around $p_y$ can be found using Eq. ({\ref{eq:probpy}) as 

\begin{equation}
P(p_y) = \frac{a}{2\pi}\left[ \frac{\sin \left(\frac{ap_y}{2\hbar}\right)}{\left(\frac{ap_y}{2\hbar}\right)}\right]^2 . \label{eq:probpy2}
\end{equation}
This is valid at the moment of passage of the particle through the slit. Marcella substitutes $p_y=p\sin \theta$ in the above expression, where $\theta$ represents the scattering angle from the centre of the slit. Putting $\alpha =ap_y/2\hbar$, one writes the above equation as

\begin{equation}
P(\alpha)=\frac{a}{2\pi}\left( \frac{\sin \alpha}{\alpha}\right)^2 \label{eq:std_fraun}
\end{equation}
 This then appears to give the standard result for Fraunhofer diffraction. But as noted by Rothman and Boughn \cite{rothman}, Marcella simply describes the state of the particle at the slit and is not concerned with what takes place at the distant observation screen.   Moreover, these authors have pointed out that the  method used in \cite{marcella} implicitly makes the same approximations as in the treatment of interference, etc.,  in classical optics. Eq. (\ref{eq:sync}) gives the probability amplitude for the momentum $p_y$ at $t=0$, which is the  Fourier transform of  the reduced wave function (\ref{eq:slit_wavefn1}), and both of them describe the same state at the  moment of detection. It is pointed out in \cite{rothman} that   Eq. (\ref{eq:std_fraun}) is not the same as the probability amplitude for the angle $\theta$ in a diffraction experiment without some  form of a `hidden variable' approach. 

In the  recent work in \cite{fabbro}, Marcella's formalism  is extended to obtain a  quantum mechanical description of diffraction through a single slit, in the Fraunhofer region and at all angular ranges.  By considering the wave function (\ref{eq:slit_wavefn1}) as a projection of an initial state $\ket{\psi_y^{in} }$ into the final one, an analysis in the light of quantum measurement theory is made. The following modifications are  made in the approach in \cite{marcella} to obtain a quantum mechanical diffraction formula: (1) The position filtering  corresponds to a measurement of the three spatial coordinates, instead of one, and  (2) it must be completed by an "energy-momentum filtering". These, it is argued, are necessary to obtain a final state compatible with a `kinematic constraint' and also with the constraint that the presence of the diffracted wave is only beyond the diaphragm. The modified theory attempts to  provide a formula for the intensity of the diffracted wave over the whole range of diffraction angle, for the case of diffraction at infinity (Frauhofer region).  However, the conceptual difficulties raised by  \cite{rothman}  are not addressed any further in \cite{fabbro}.

\section{Location state: free particle} \label{sec:free}

Before going into the theory of diffraction, we consider  the measurement of position of  a particle in one dimension.   In the standard framework of quantum mechanics,  if the system consists of a single particle in one-dimension (whose position coordinate is denoted here as $y$), and if the measurement is  of this observable $y$ with infinite precision at some point $y^{\prime}$,  the initial state gets reduced to a position eigenstate $\ket{y^{\prime}}$. In the position representation, this  eigenket  can be written as \cite{wise,sakurai}

\begin{equation}
\braket{y|y^{\prime}} =\delta (y-y^{\prime}) ,
\end{equation}
which is the Dirac $\delta$-function. One can write this as

\begin{equation}
\braket{y|y^{\prime}} =\delta (y-y^{\prime}) = \frac{1}{2\pi} \int_{-\infty}^{\infty}dk_y\; e^{ik_y(y-y^{\prime})} . \label{eq:closure1}
\end{equation}
It may be noted that this is the same as the closure relation in this case.

 On the other hand, in  an actual position measurement such as the one described in the above section,    the  collapsed wave function is postulated to take the form (\ref{eq:slit_wavefn1}). But this can be written as a superposition of  position eigenkets, 

\begin{equation}
\psi_y^{0,a}(y) = \frac{1}{ \sqrt{a}}\int ^{a/2}_{-a/2} dy^{\prime} \;\delta (y-y^{\prime}) . \label{eq:slit_wavefn2}
\end{equation} 
Using equation (\ref{eq:closure1}), this expression can be seen to be

\begin{equation}
\psi_y^{0,a}(y)=\frac{1}{2\pi \sqrt{a}}\int_{-a/2}^{a/2}dy^{\prime}\int_{-k_{m}}^{k_{m}}dk_y\; e^{ik_y(y-y^{\prime})}, \label{eq:slit_wavefn3}
\end{equation}
where the limit $k_{m}\rightarrow \infty$ has to be taken. One can rewrite this equation as 

\begin{equation}
\psi_y^{0,a}(y)=\frac{1}{2\pi \sqrt{a}}\int_{-k_{m}}^{k_{m}}dk_y \; \left[ \int_{-a/2}^{a/2}dy^{\prime} e^{-ik_y y^{\prime}} \right]   e^{ik_y y}, \label{eq:slit_wavefn3a}
\end{equation}
which is a superposition of  eigenstates  $e^{ik_yy}$ of the Hamiltonian operator in this case, with the coefficients contained in the square bracket. Let  this  wave function be denoted  as $\Psi_y(y,0)\equiv \psi_y^{0,a}(y)$. We can introduce the unitary time evolution of the wave function for $t>0$  as \cite{gottfried}
 
 \begin{equation}
\Psi_y(y,t)=\frac{1}{2\pi \sqrt{a}}\int_{-k_{m}}^{k_{m}}dk_y \; \left[ \int_{-a/2}^{a/2}dy^{\prime} e^{-ik_y y^{\prime}} \right]   e^{ik_y y}\; e^{-iE_{y} t/\hbar}, \label{eq:slit_wavefn4}
\end{equation}
 under the limit $k_m \rightarrow \infty$. Here $E_{y}=\hbar^2k_y^2/2m$. In this form, we call it the `position measurement-induced collapse state' or `location state'. It is important to note that  such time-evolution of the wave function is not considered in \cite{marcella}.

\begin{figure}[h]
\centering 
\resizebox {0.3 \textwidth} {0.20 \textheight}{\includegraphics{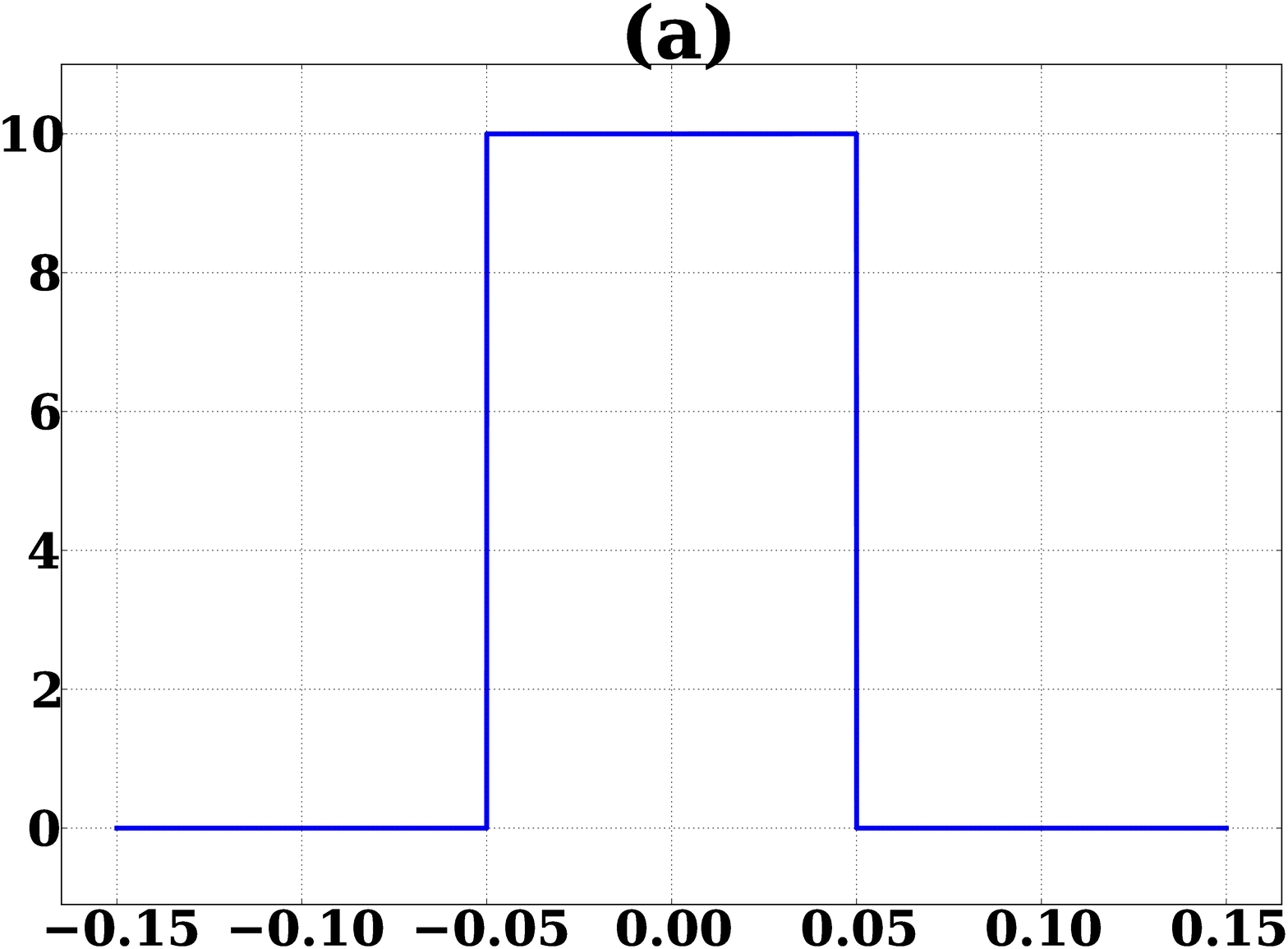}} 
\resizebox {0.3 \textwidth} {0.20 \textheight }{\includegraphics{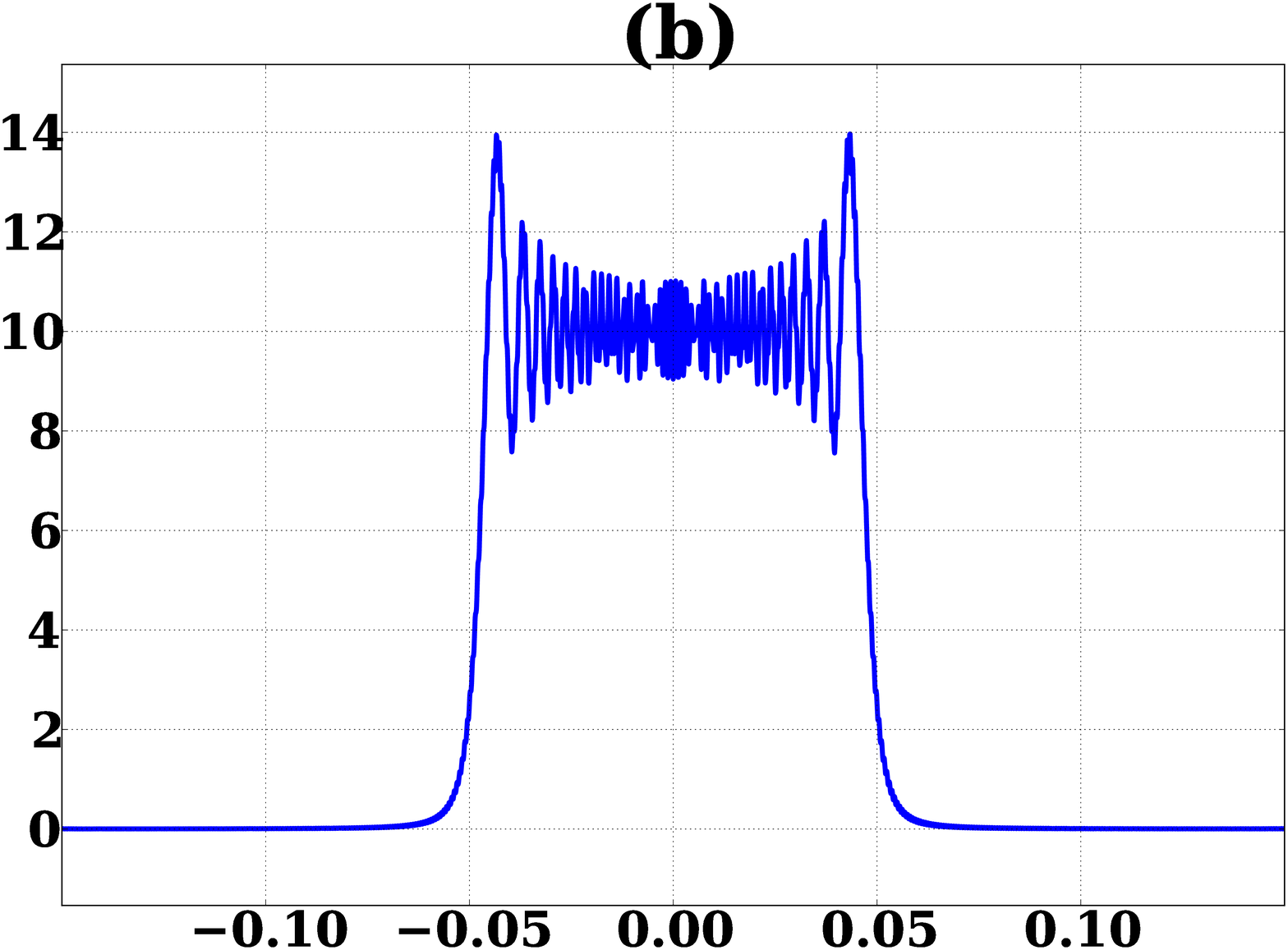}} 
\resizebox {0.3 \textwidth} {0.20 \textheight }{\includegraphics{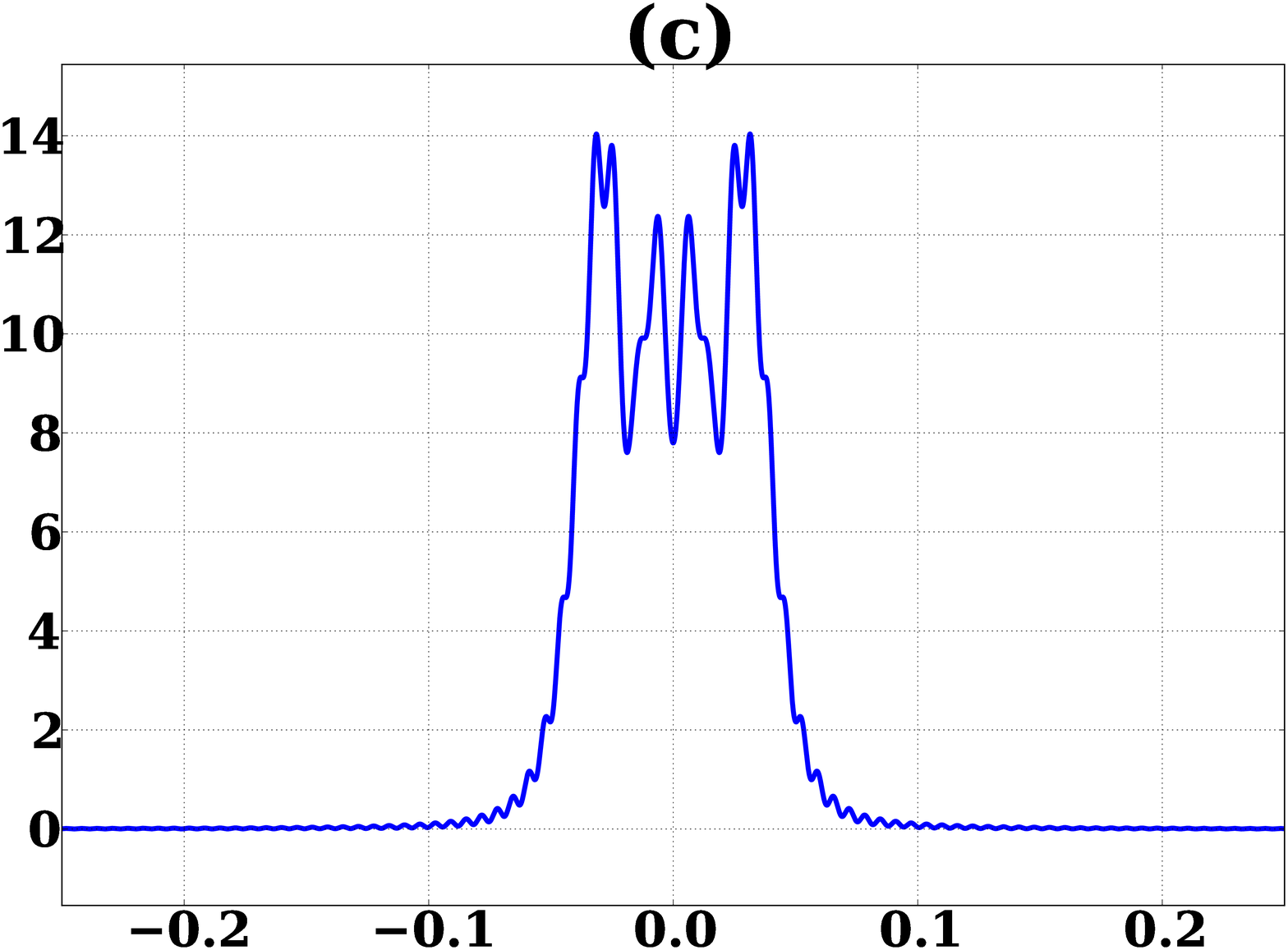} } 
\resizebox {0.3 \textwidth} {0.20 \textheight }{\includegraphics{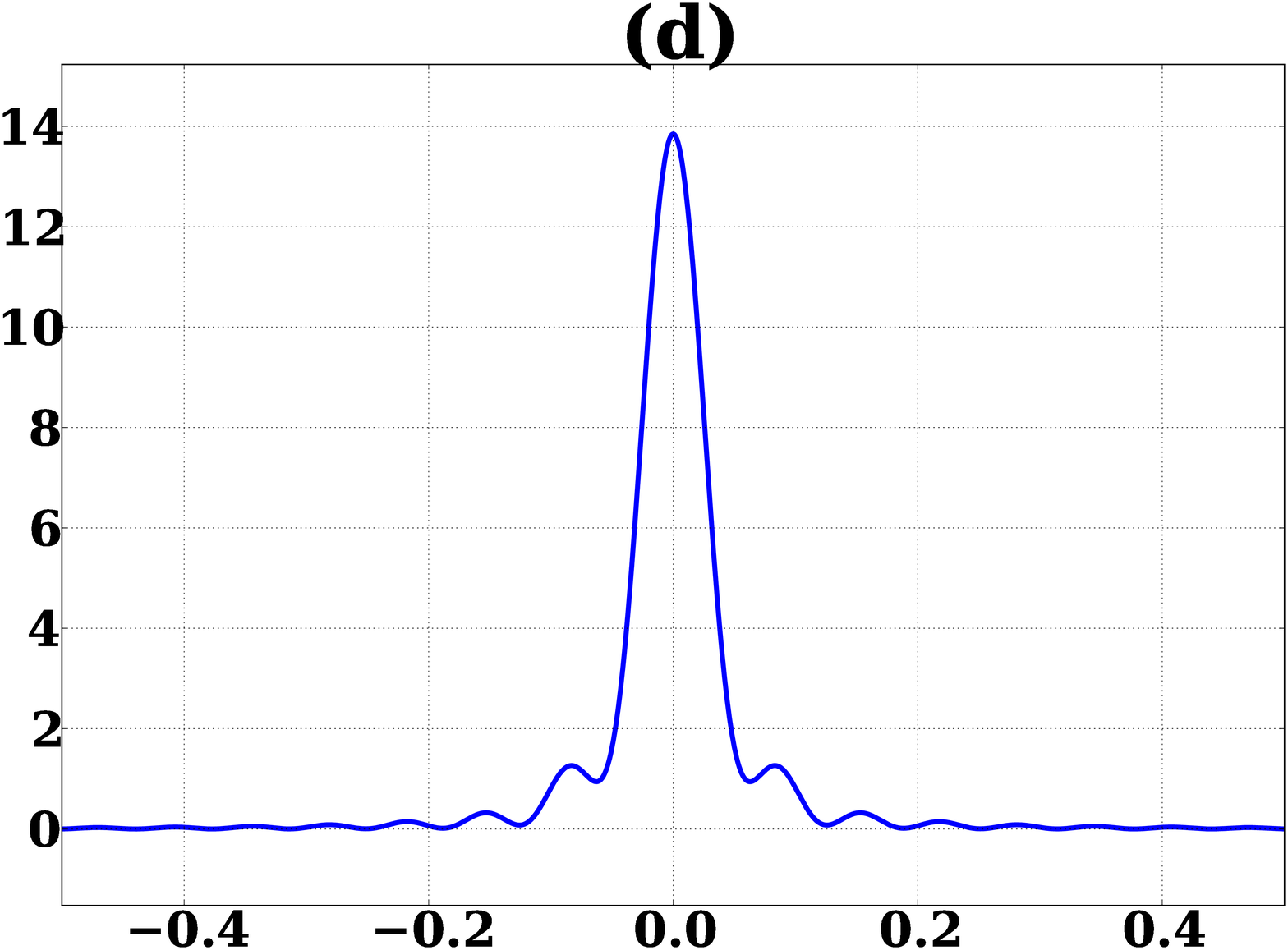}}
 \resizebox {0.3 \textwidth} {0.20 \textheight }{\includegraphics{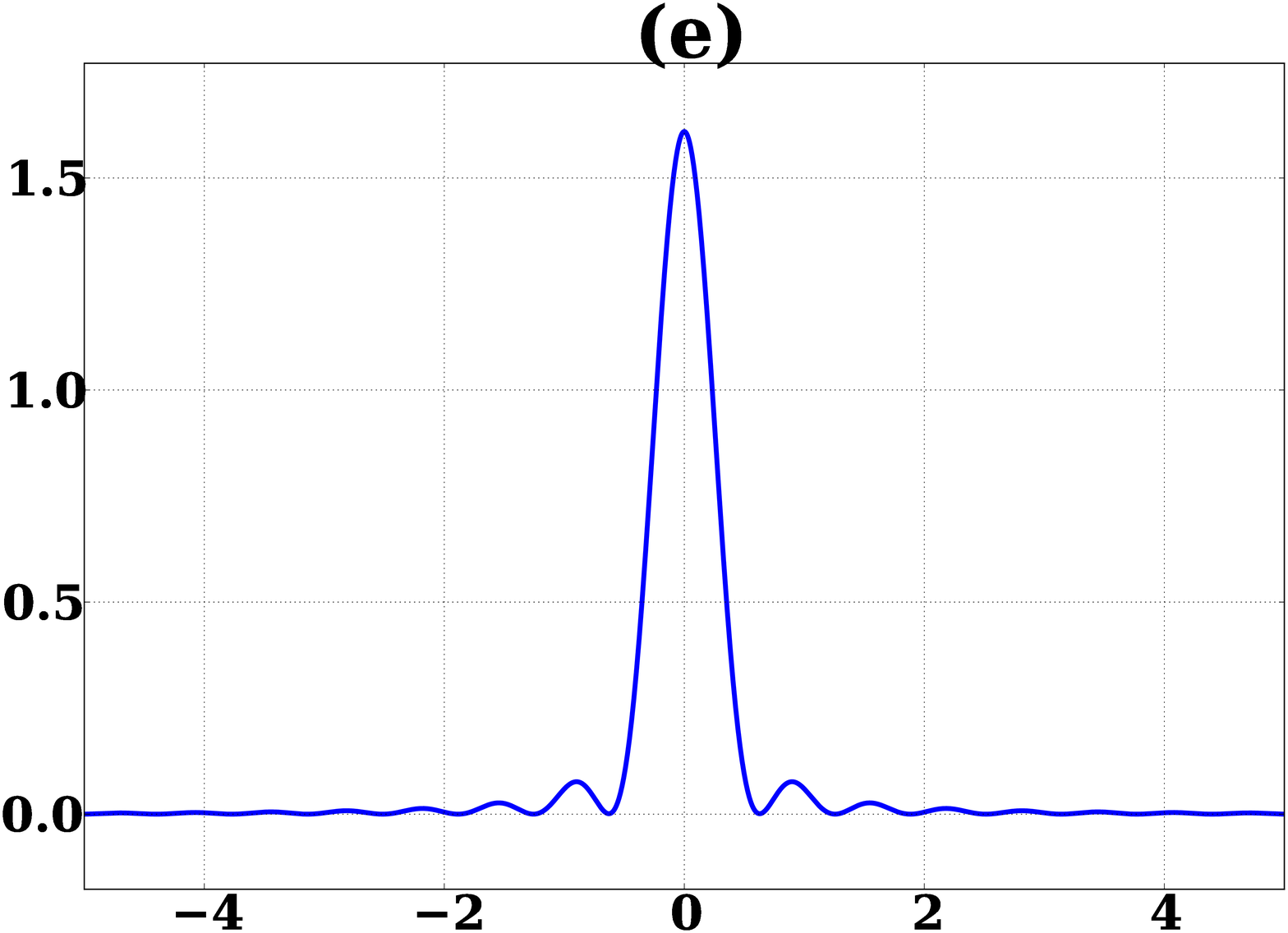}} 
 \resizebox {0.3 \textwidth} {0.20 \textheight }{\includegraphics{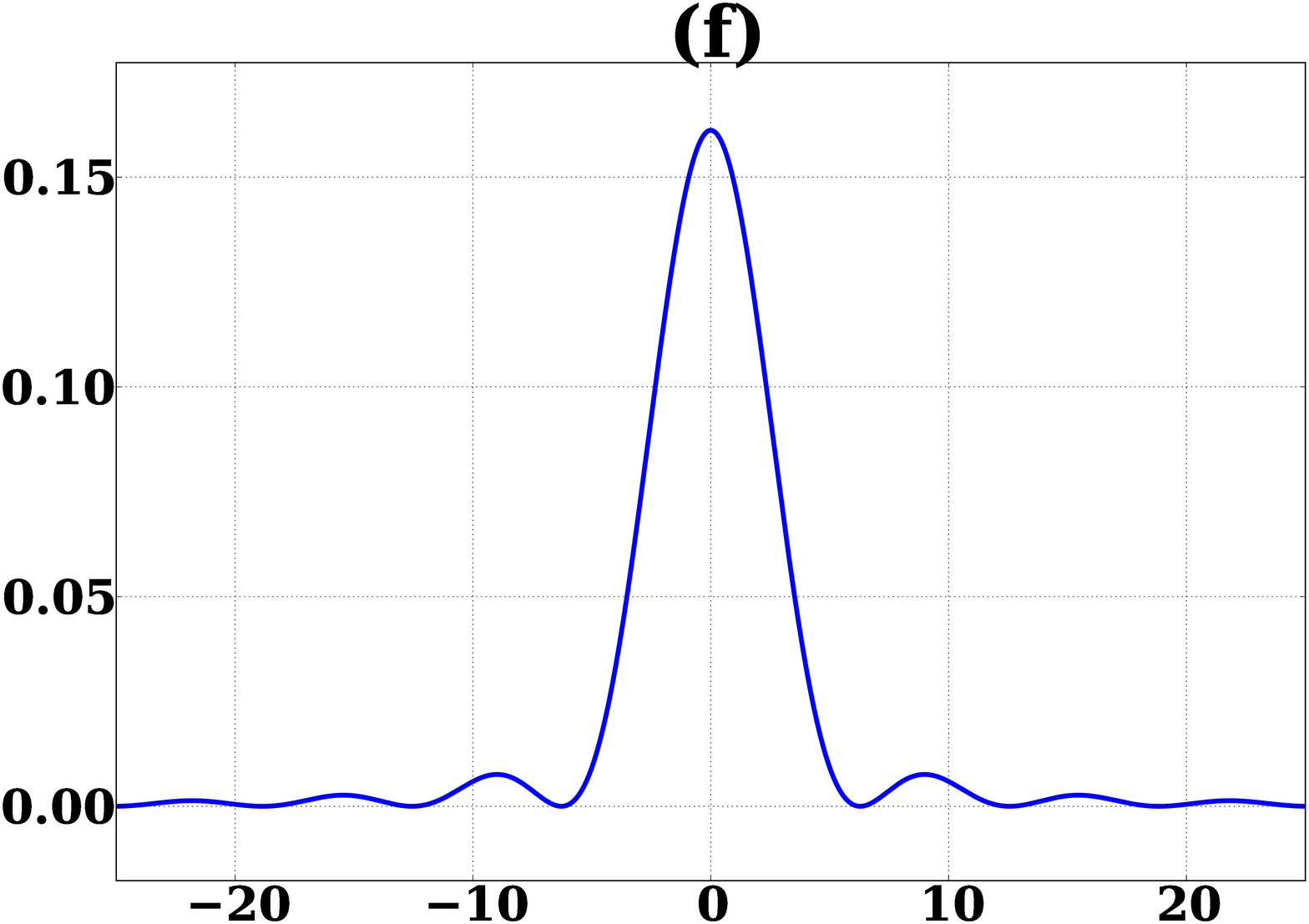}}
\caption{Probability distribution for the location state of a free particle in  one-dimension with time $T$= (a) $0$, (b) $10^{-5}$, (c) $10^{-4}$, (d) $10^{-3}$, (e) $10^{-2}$, (f) $10^{-1}$}
\label{fig:pmic_free}        
\end{figure}

 We have plotted this   location state wave function in Eq. (\ref{eq:slit_wavefn4}) for various values of $t$, putting $\hbar/m=1$, $a=0.1$ and  by fixing $k_m=10^{8}$, in Fig. \ref{fig:pmic_free}. This value of $k_m$  appears to be reasonably large, for any further increase in it does not lead to significant change in the plot of $|\Psi_y(y,t)|^2$ versus $y$, for all values of $t$.  For $t=0$, we are able to regain  the  real rectangular wave function given in Eq. (\ref{eq:slit_wavefn1}) with very good accuracy. Thus without the factor $e^{-iE_{y} t/\hbar}$ in the integral, this gives nothing new, except the original reduced wave function.  It is notable that this commonly conceived  wave function in Eq. (\ref{eq:slit_wavefn1}) has a spreading for $t>0$. As can be seen from the figures, it is not simply a  spreading of the rectangular wave function in (\ref{eq:slit_wavefn1}) as such; for small values of $t$, the modulus square of the wave function in position space acquires the shape of the intensity distribution for Fresnel diffraction and for large $t$, it approaches the pattern for Fraunhofer diffraction.

 In summary,  our method consists of first rewriting the rectangular  wave function into the form (\ref{eq:slit_wavefn2}), then  substituting for the $\delta$-function in the integrand with the right hand side of Eq. (\ref{eq:closure1}) (which is the closure relation in this case), and finally  postulating the unitary time-evolution as per Eq. (\ref{eq:slit_wavefn4}). Note that in this section, we have only described the spreading of a one-dimensional rectangular wave function with time and have plotted the probability density, with the results resembling the   diffraction patterns. For an actual treatment of diffraction, we must specify the value of time corresponding to the distance to the screen, which we shall do in Sec. \ref{sec:diffraction}.

To our knowledge, the above method of obtaining the patterns similar to Fresnel and Fraunhofer from the spreading of the same rectangular wave function is not discussed elsewhere in the literature.  This formalism to describe the time-evolution of a rectangular wave function  shall be useful not only for diffraction, but also for other cases of position measurement of particles moving in general potentials, as  demonstrated in the next section.

\section{Location state: general potential} \label{sec:general}

Let us now consider the  general case of position measurement of a particle  in a potential $V(y)$, where the  measurement is  made at some arbitrary point $y_0$. Here we assume that the collapsed wave function at $t=0$ is nonzero only between $y_0 -a/2$ and $y_0+a/2$. In the simpler case,  this state is described by a constant wave function similar to that in Eq. (\ref{eq:slit_wavefn1}), but centred at $y_0$,  at the moment of measurement. Let the eigenkets of the Hamiltonian operator in this case be denoted by $\ket{u_n}$, so that the corresponding position space energy eigenfunctions are $u_n(y)\equiv \braket{y|u_n}$. If we  perform  an idealised measurement  at the point $y^\prime$ with  infinite precision, the resulting reduced wave function would be  a $\delta$-function. One can  use  the closure relation

\begin{equation}
\sum_n u_n^{\star}(y^{\prime}) u_n(y) = \delta (y-y^{\prime}), \label{eq:closure}
\end{equation}
to represent this position eigenstate while it is in this potential. This can be used  to expand the reduced wave function  (\ref{eq:slit_wavefn1}) at $t=0$, as in Eq. (\ref{eq:slit_wavefn2}), 

\begin{equation}
\psi_y^{y_0,a}(y)=\frac{1}{2\pi \sqrt{a}}\int_{y_0-a/2}^{y_0+a/2}dy^{\prime}\sum_n u_n^{\star}(y^{\prime}) u_n(y), \label{eq:slit_wavefn5}
\end{equation}
where  the limit $n\rightarrow \infty$ must be taken. We  denote this wave function as $\Psi_y(y,0)\equiv \psi_y^{y_0,a}(y) $.  As in the previous case, we rewrite the above equation and introduce the time-evolution of the wave function of the particle   as 

\begin{equation}
\Psi_y(y,t)=\frac{1}{2\pi \sqrt{a}} \sum_n  \left[ \int_{y_0-a/2}^{y_0+a/2}dy^{\prime}\;u_n^{\star}(y^{\prime}) \right]u_n(y)e^{-iE_nt/\hbar},  \label{eq:slit_wavefn6}
\end{equation}
where again the limit $n\rightarrow \infty$ may be taken. Here $E_n$ are the energy eigenvalues of the particle when it is in this potential.

\begin{figure}[h]
\centering 
\resizebox {0.3 \textwidth} {0.20 \textheight }{\includegraphics {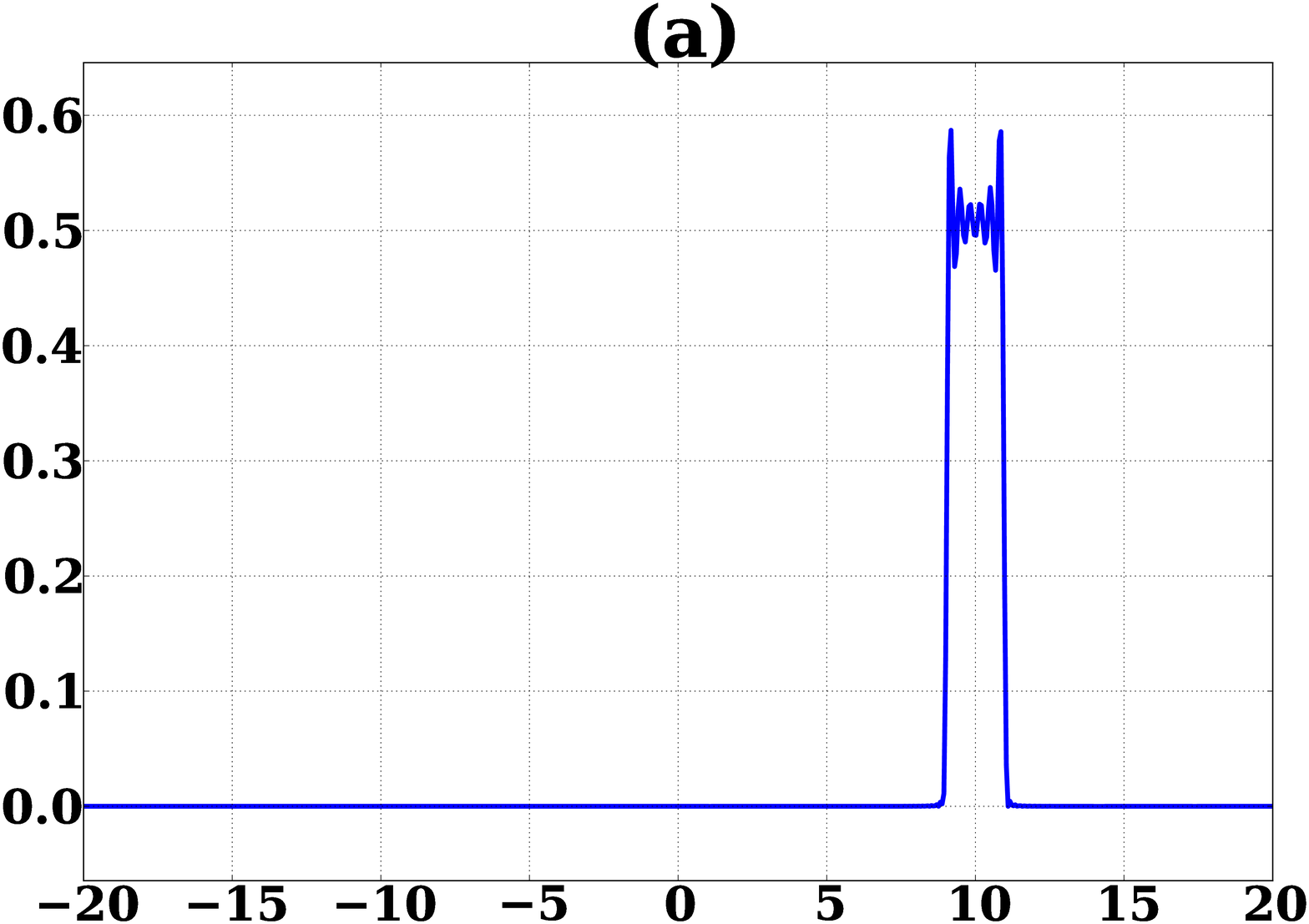}}
\resizebox {0.3 \textwidth} {0.20 \textheight }{\includegraphics {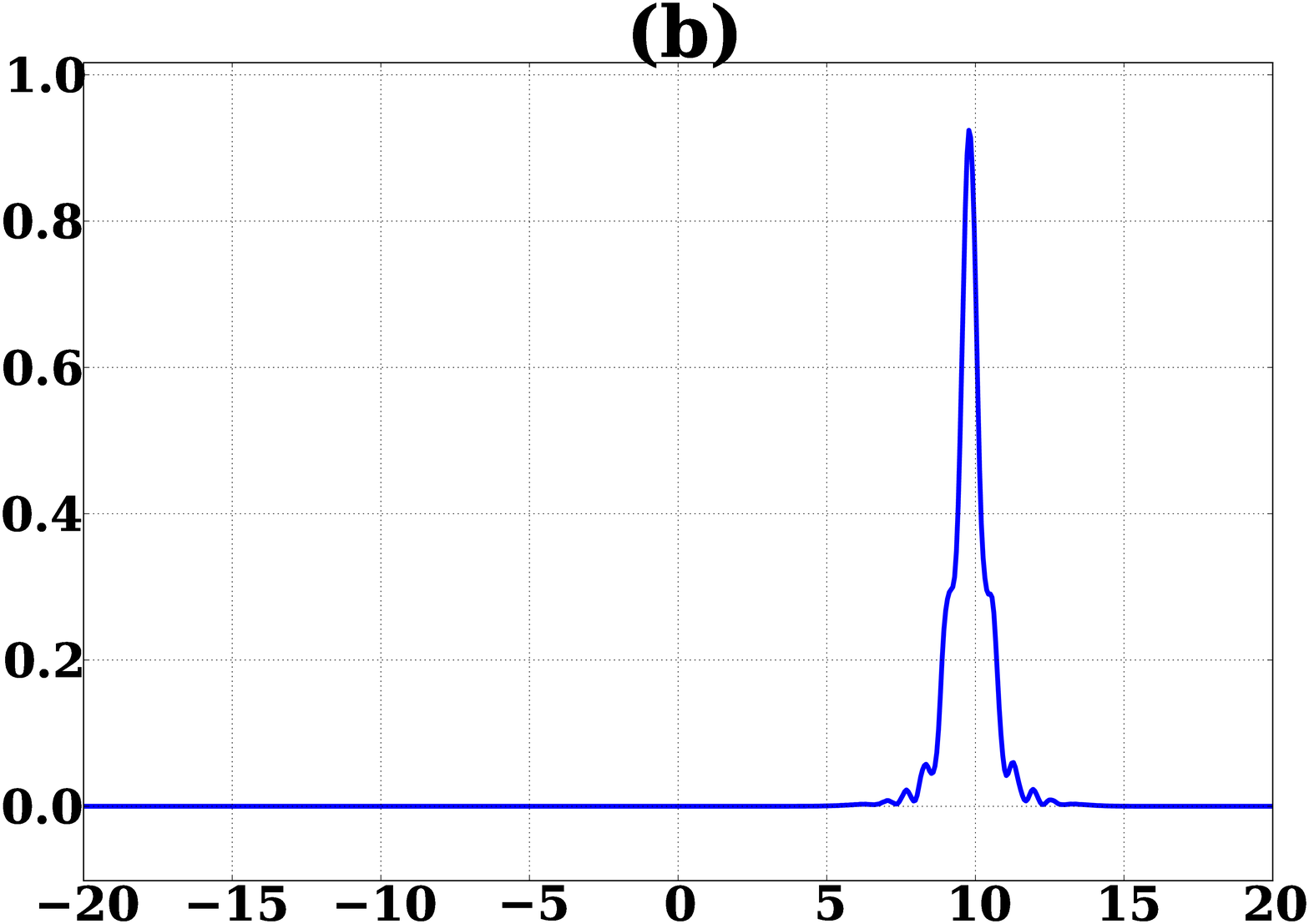}}
\resizebox {0.3 \textwidth} {0.20 \textheight }{\includegraphics {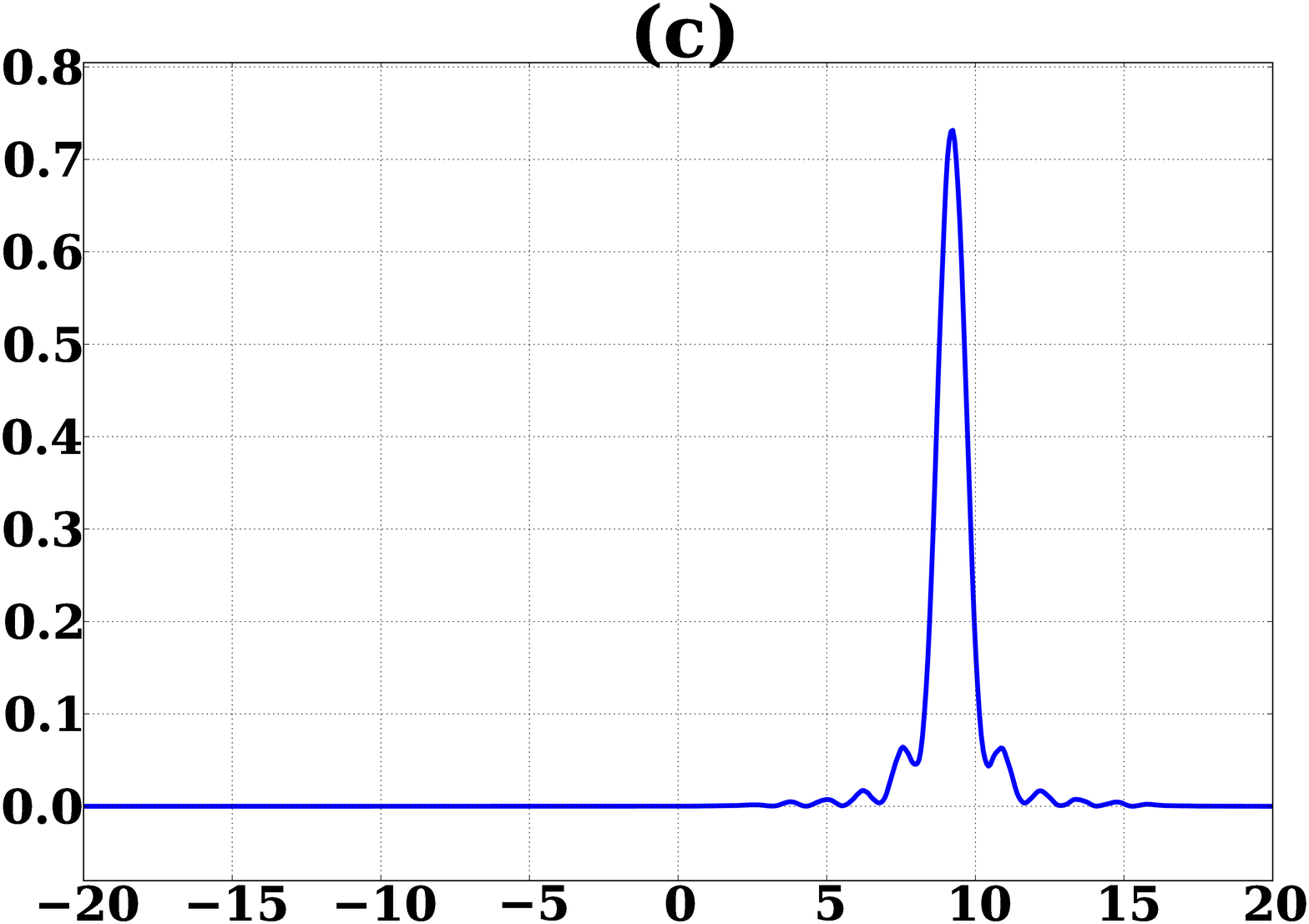}}
\resizebox {0.3 \textwidth} {0.20 \textheight }{\includegraphics {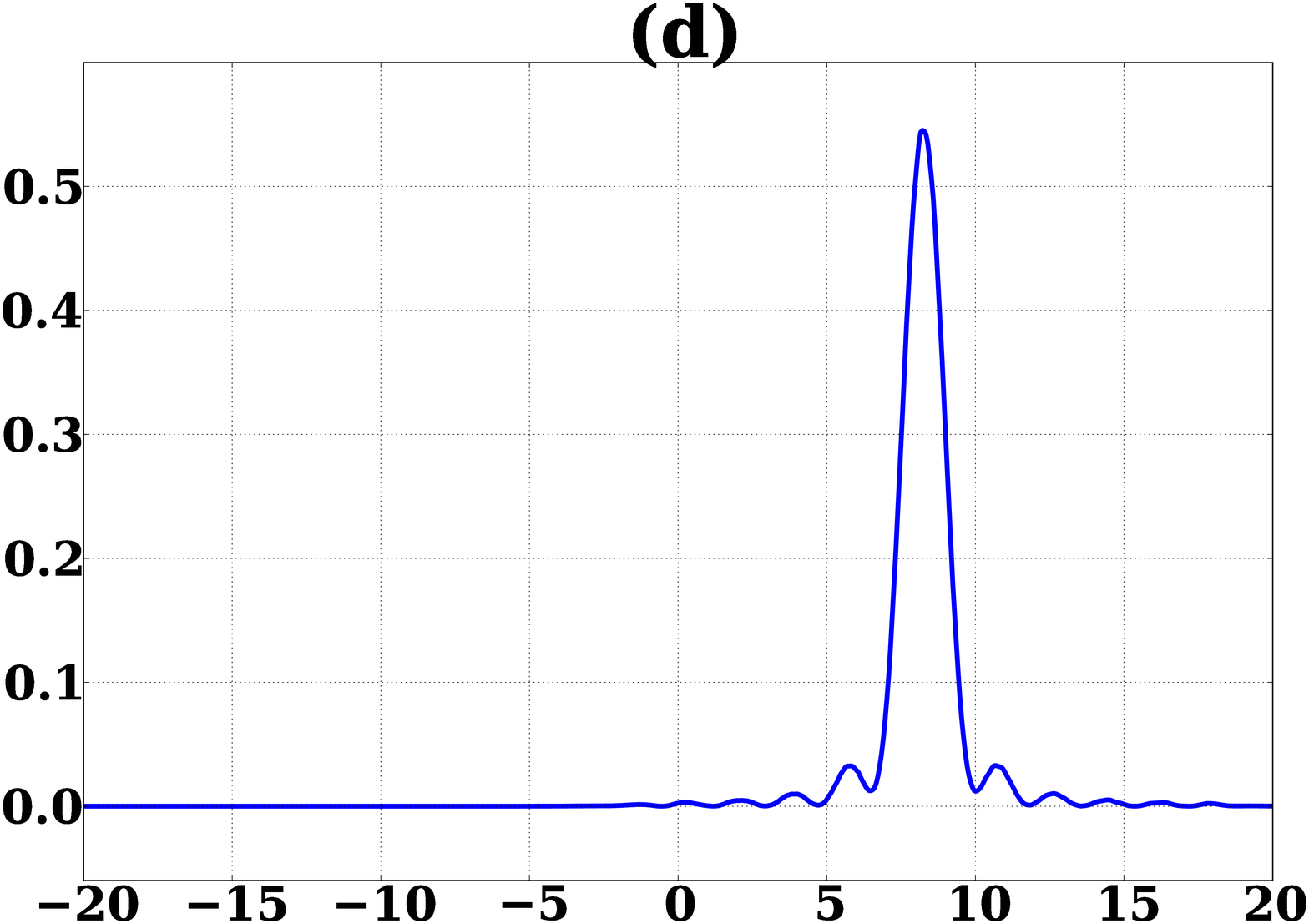}}
\resizebox {0.3 \textwidth} {0.20 \textheight }{\includegraphics {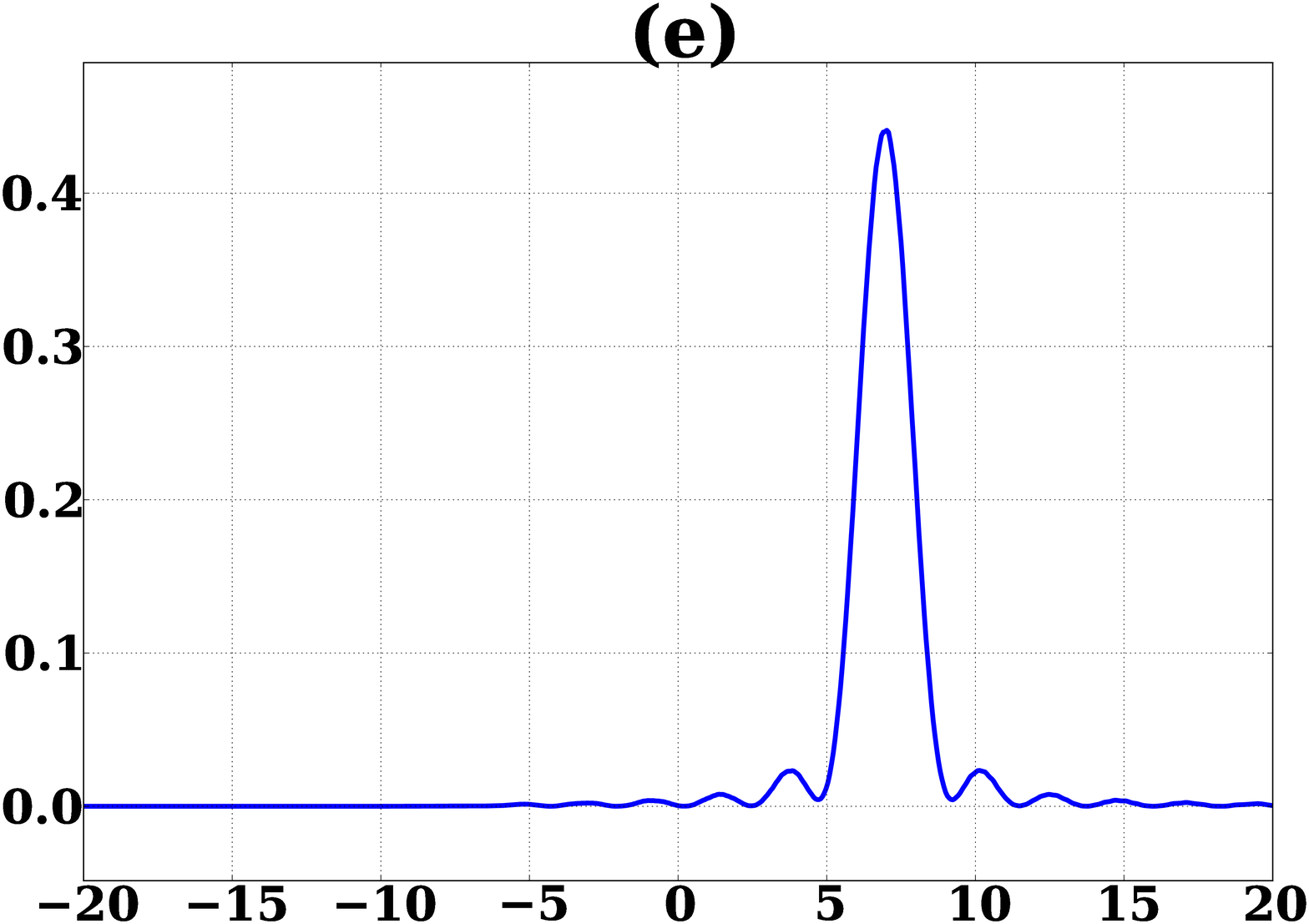}}
\resizebox {0.3 \textwidth} {0.20 \textheight }{\includegraphics {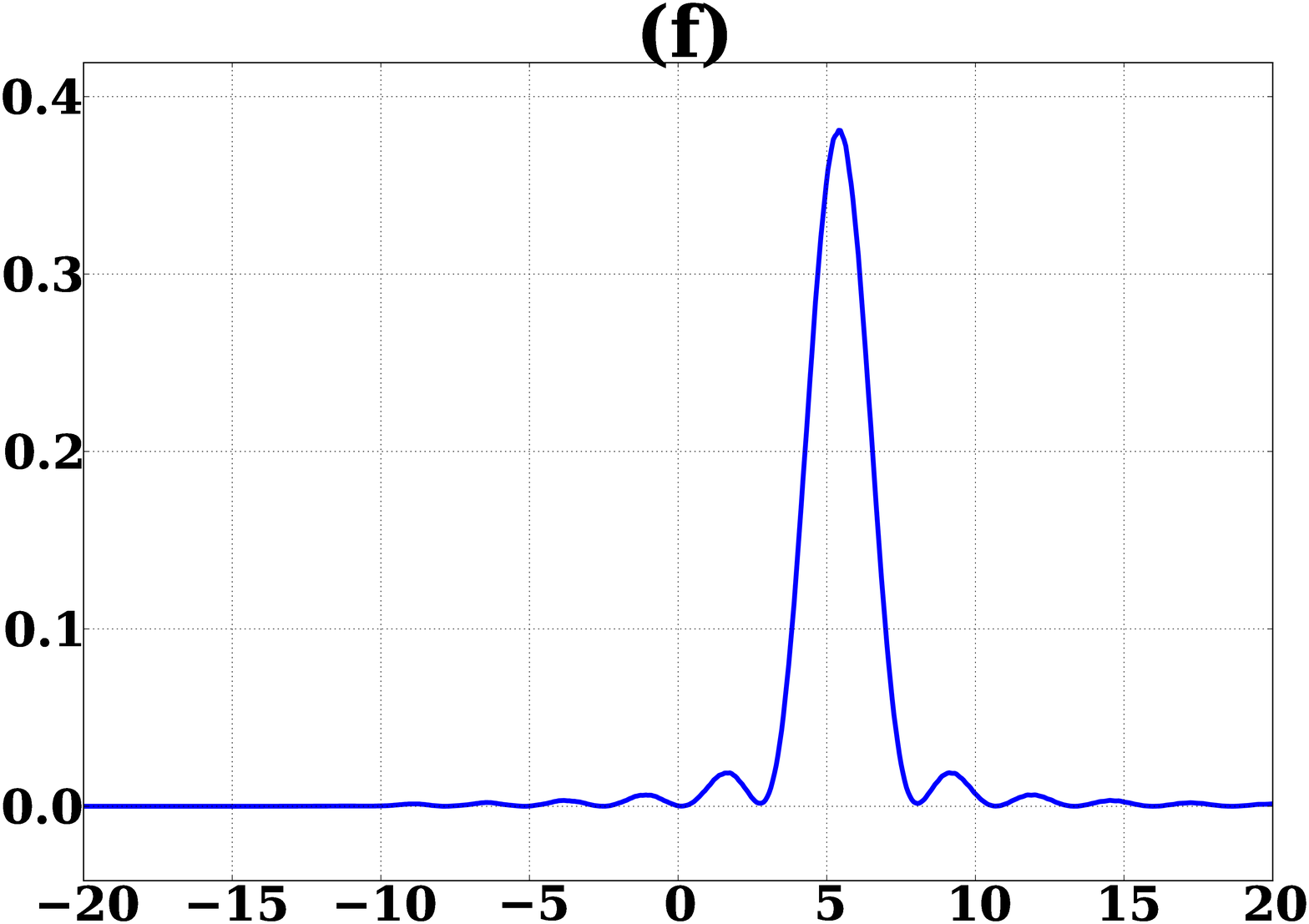}}
\caption{Probability distribution for location states of a harmonic oscillator for $T$= (a) 0.0, (b) 0.2, (c) 0.4, (d) 0.6, (e) 0.8 (f) 1.0.}
\label{fig:pmic_shm}
 \end{figure}  

We have plotted the above wave function for a harmonic oscillator with potential $V(y)=\frac{1}{2}m\omega^2 y^2$. Here we use

\begin{equation}
u_n(y) = (2^n n! \sqrt{\pi}\sigma)^{-1/2}\exp\left[ -\frac{1}{2}\left( \frac{y}{\sigma}\right)^2\right] H_n\left(\frac{y}{\sigma}\right)
\end{equation}
and the corresponding energy eigenvalues $E_n=\hbar \omega (n+\frac{1}{2})$, with $\sigma=\sqrt{\hbar/(m\omega)}$ in the above equation (\ref{eq:slit_wavefn6}). The oscillator is assumed to be detected  at $y_0=10$ by a slit of width $a=2$, with $\hbar/m=1$. The plots  given   in Fig. \ref{fig:pmic_shm} are for different values of $t$, with an upper limit for $n$ as $n_{m}=250$. It can be seen that the pattern gradually changes from that of Fresnel diffraction to Fraunhofer diffraction.

\begin{figure}[h]
\centering 
\resizebox {0.23 \textwidth} {0.172 \textheight }{\includegraphics {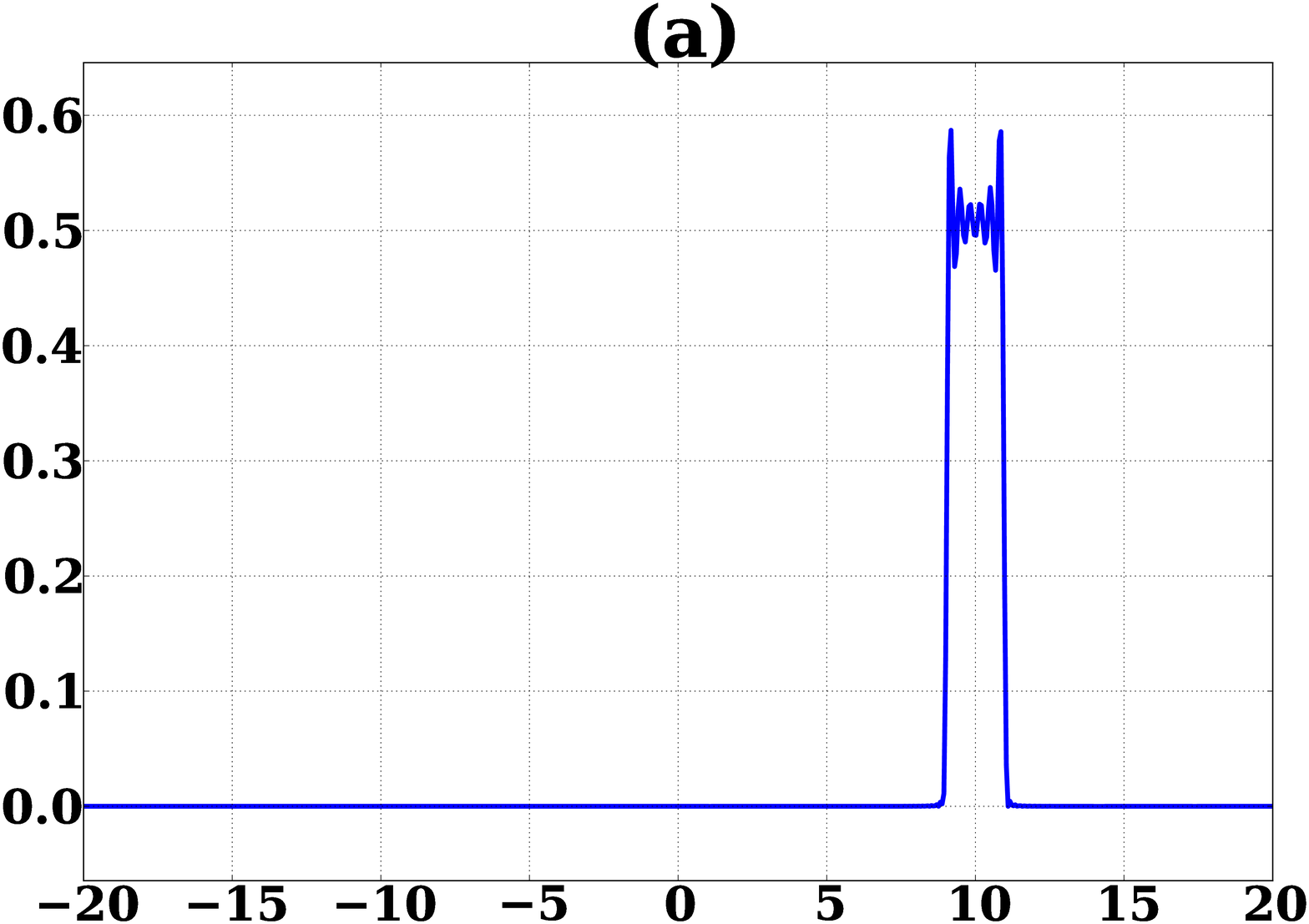}}
\resizebox {0.23 \textwidth} {0.172 \textheight }{ \includegraphics{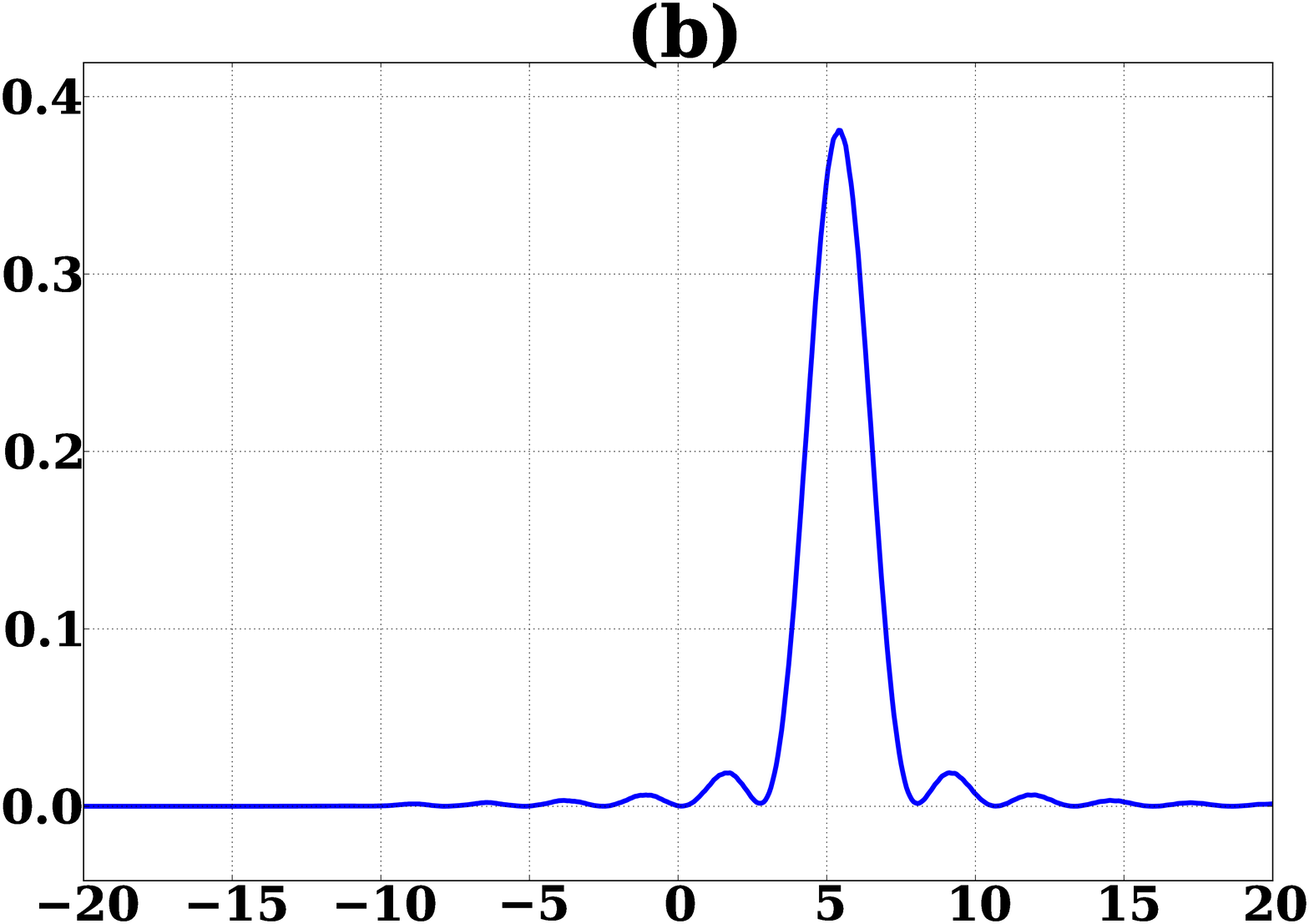}}
\resizebox {0.23 \textwidth} {0.172 \textheight }{ \includegraphics{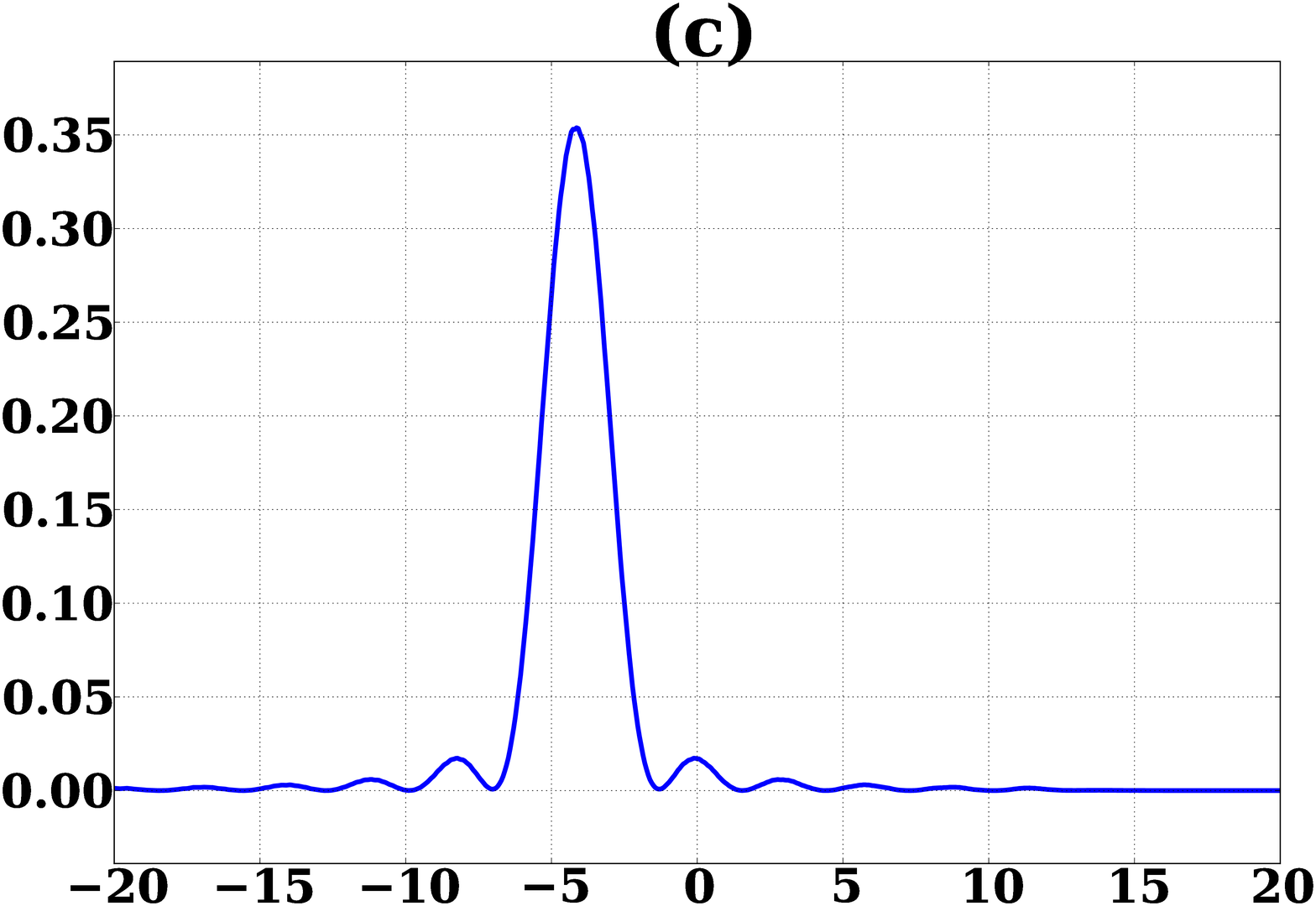}} 
\resizebox {0.23 \textwidth} {0.172 \textheight }{ \includegraphics{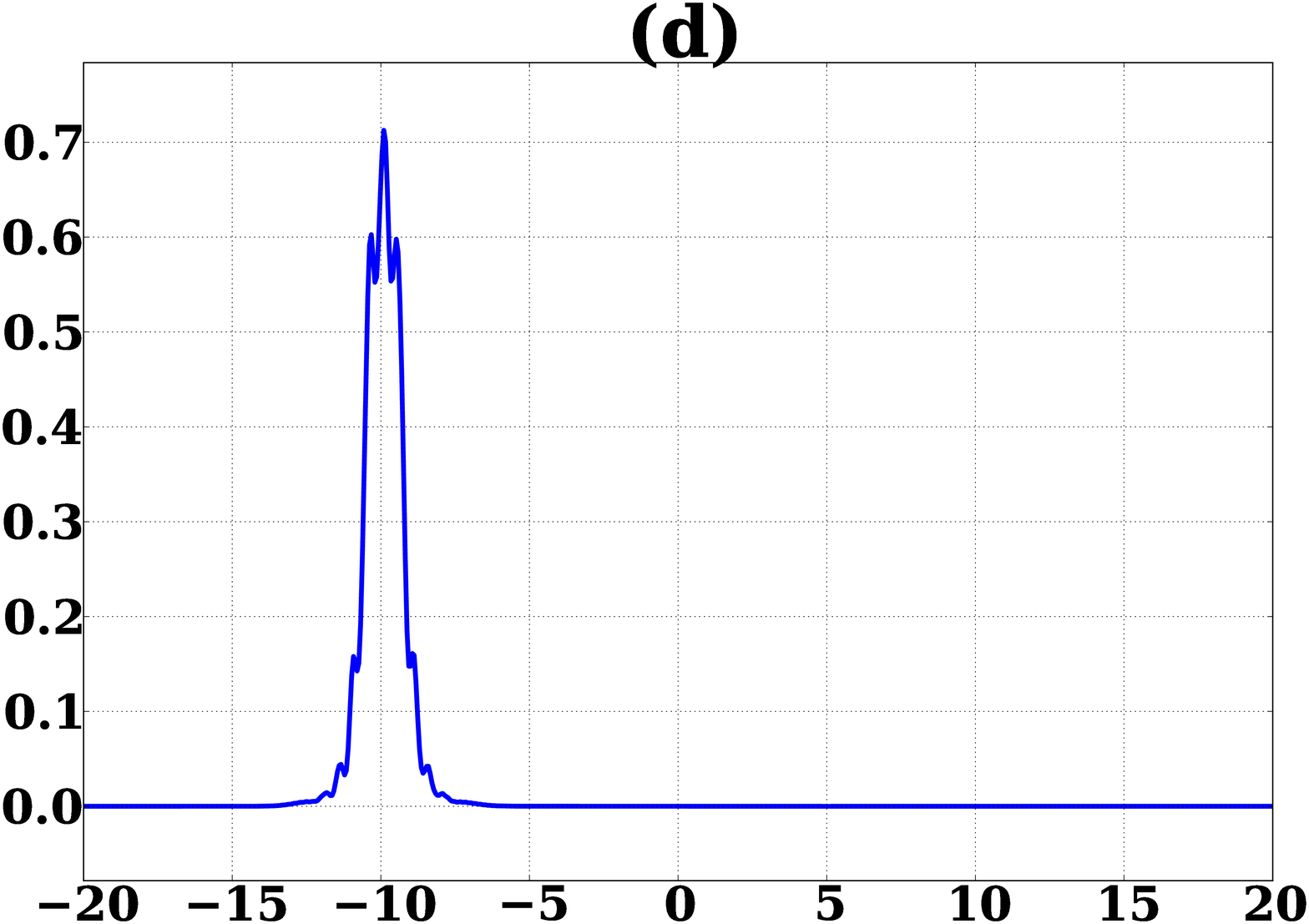} }
\resizebox {0.23 \textwidth} {0.172 \textheight }{\includegraphics{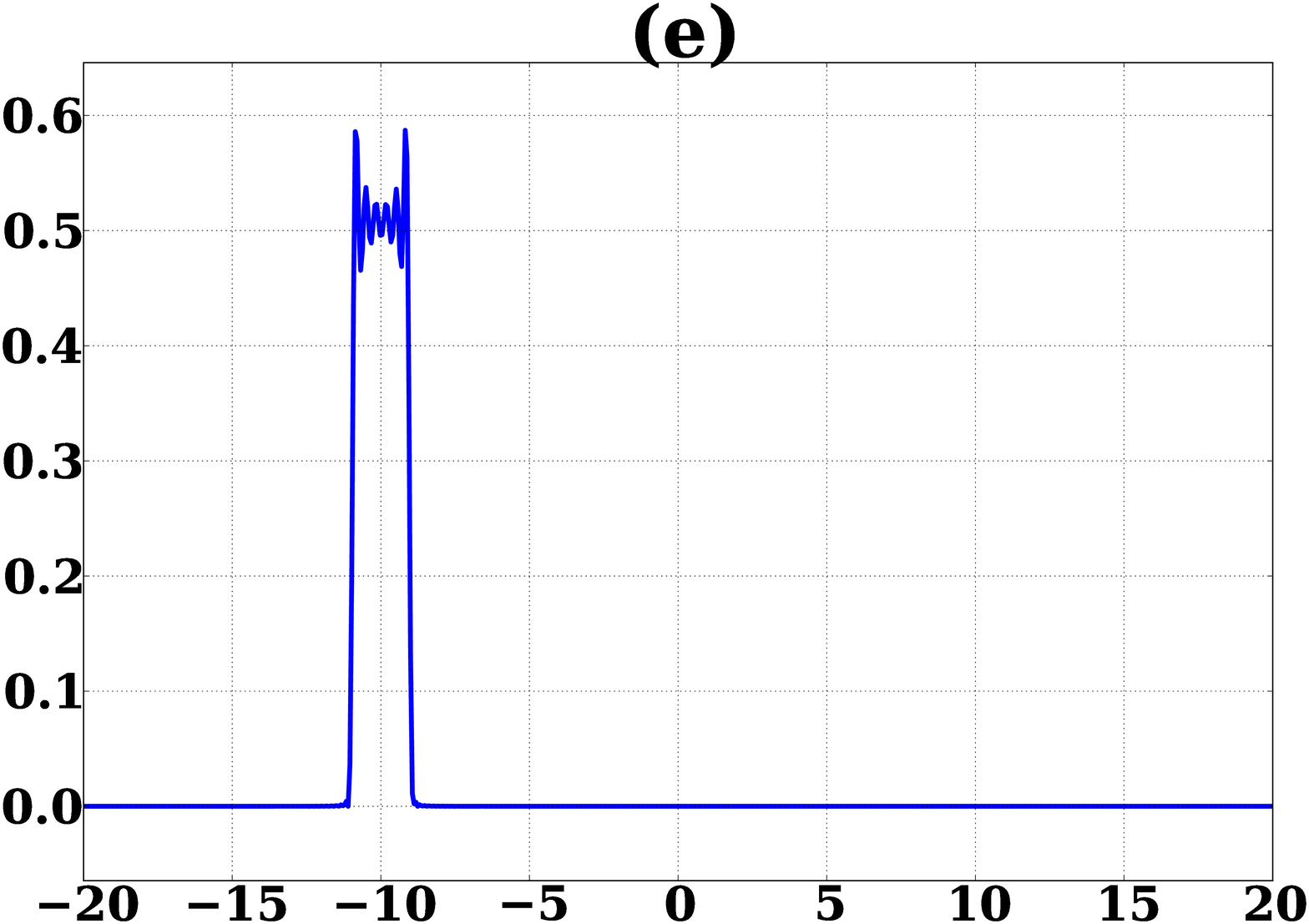}}
\resizebox {0.23 \textwidth} {0.172 \textheight }{  \includegraphics{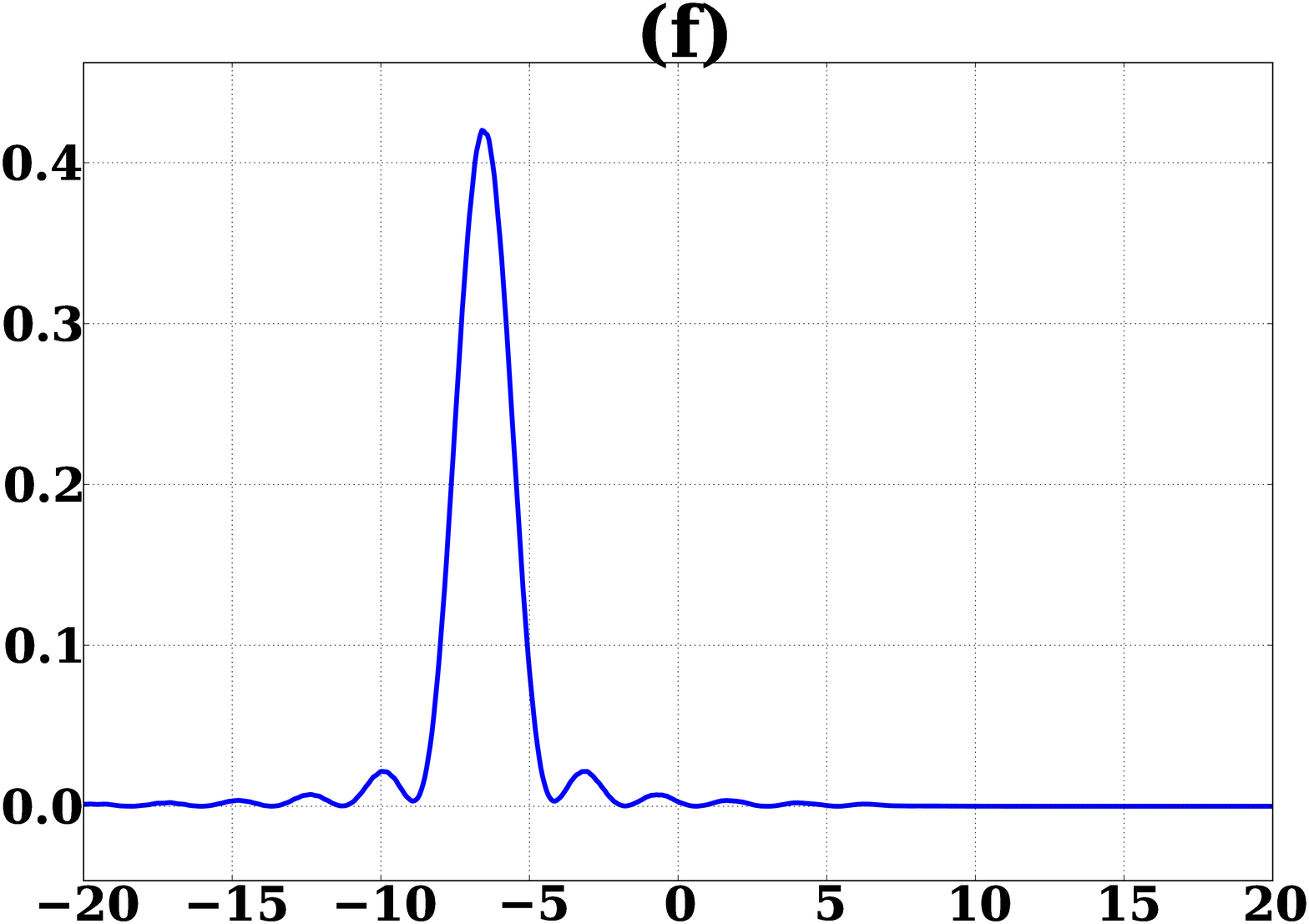} }
\resizebox {0.23 \textwidth} {0.172 \textheight }{   \includegraphics{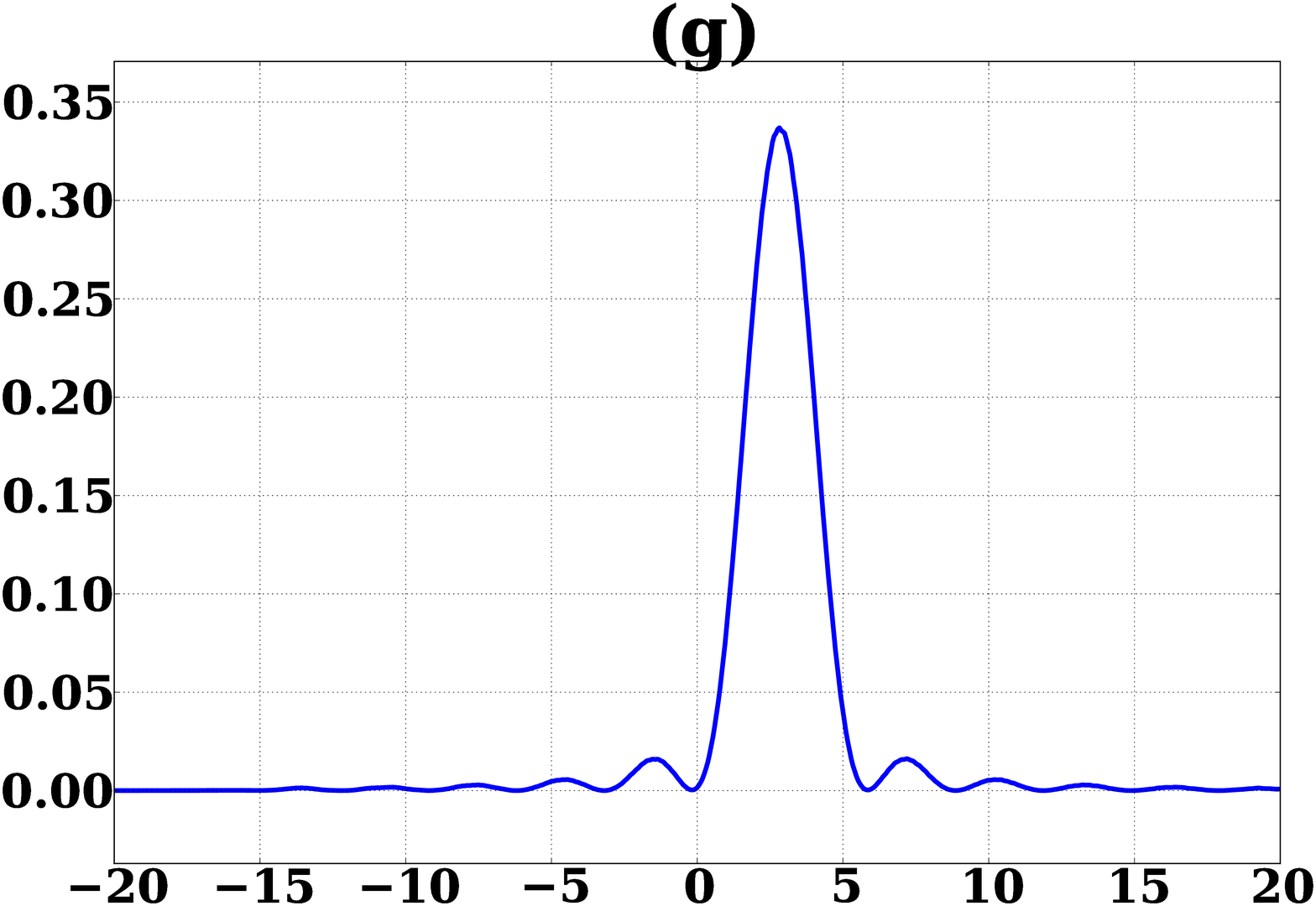} }
\resizebox {0.23 \textwidth} {0.172 \textheight }{  \includegraphics{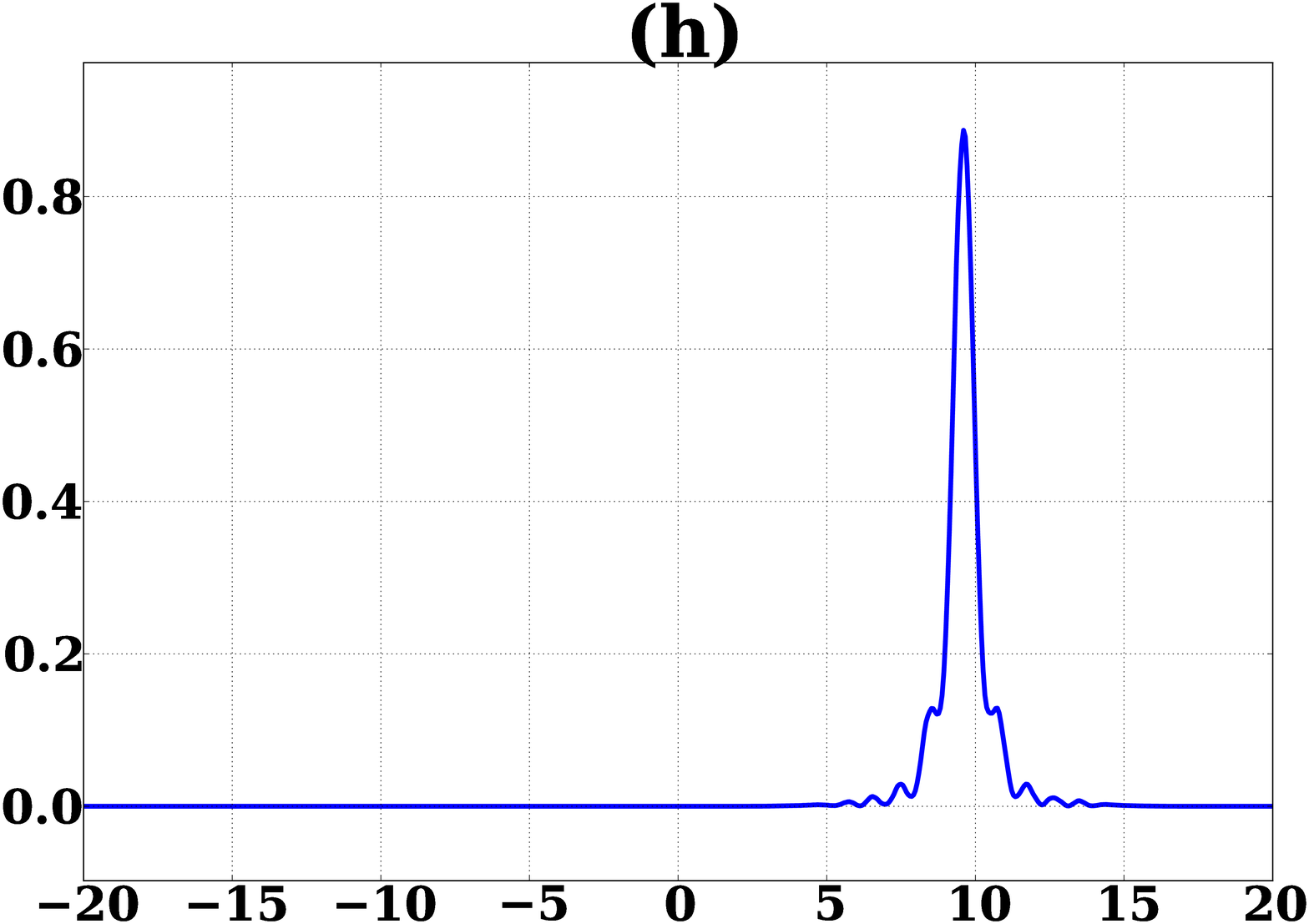}}
\resizebox {0.23 \textwidth} {0.172 \textheight }{  \includegraphics{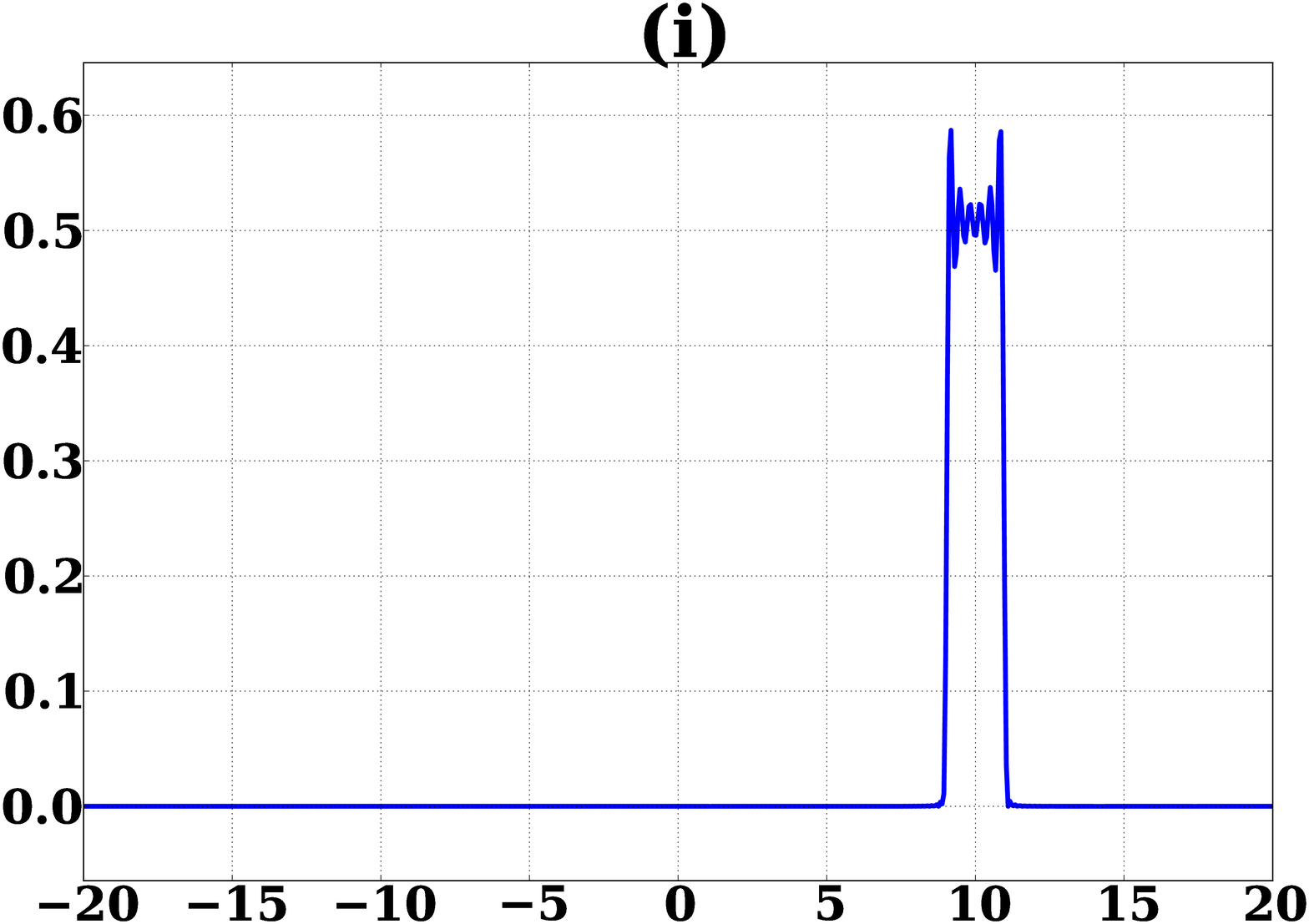}}
\resizebox {0.23 \textwidth} {0.172 \textheight }{   \includegraphics{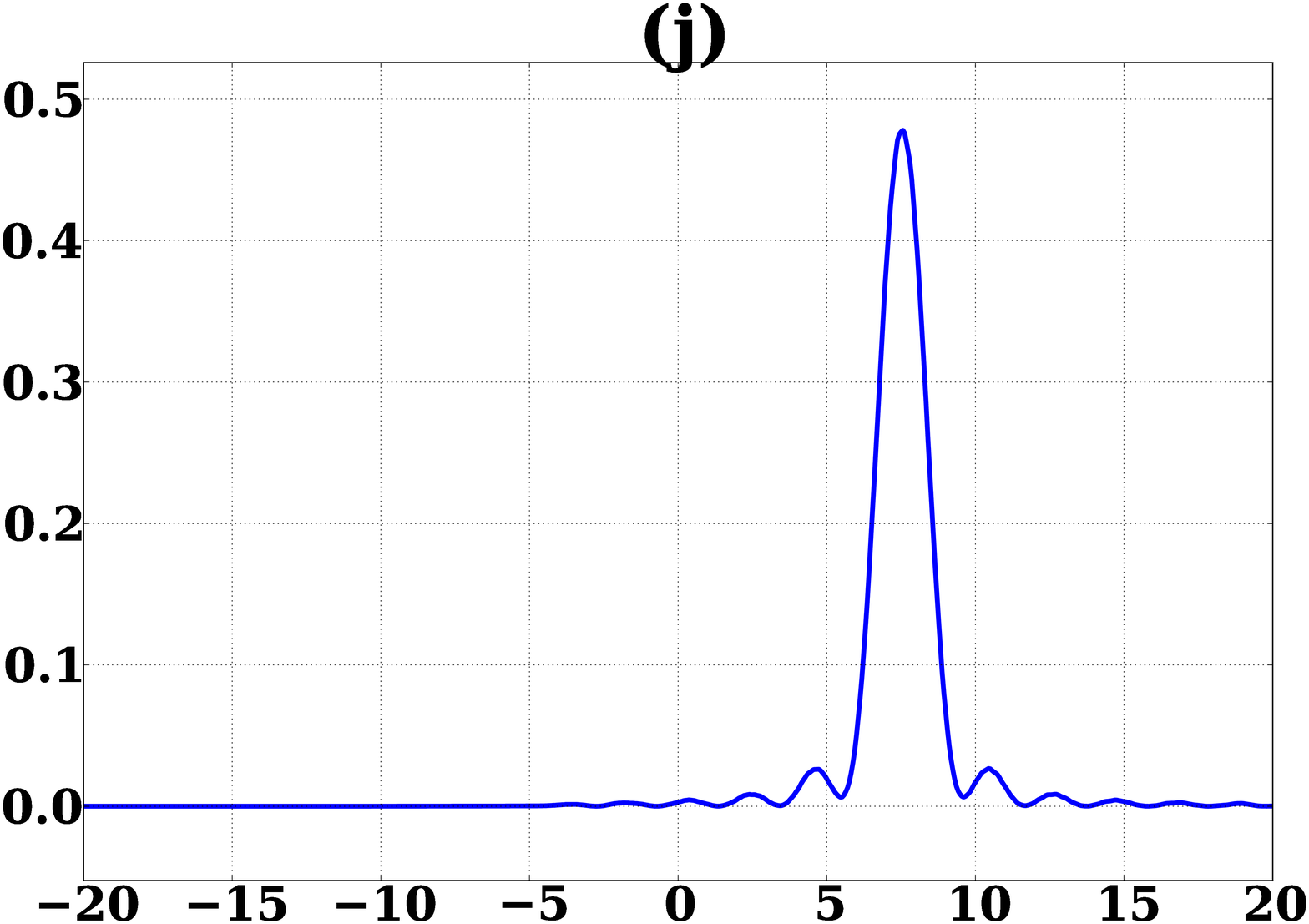}}
\resizebox {0.23 \textwidth} {0.172 \textheight }{     \includegraphics{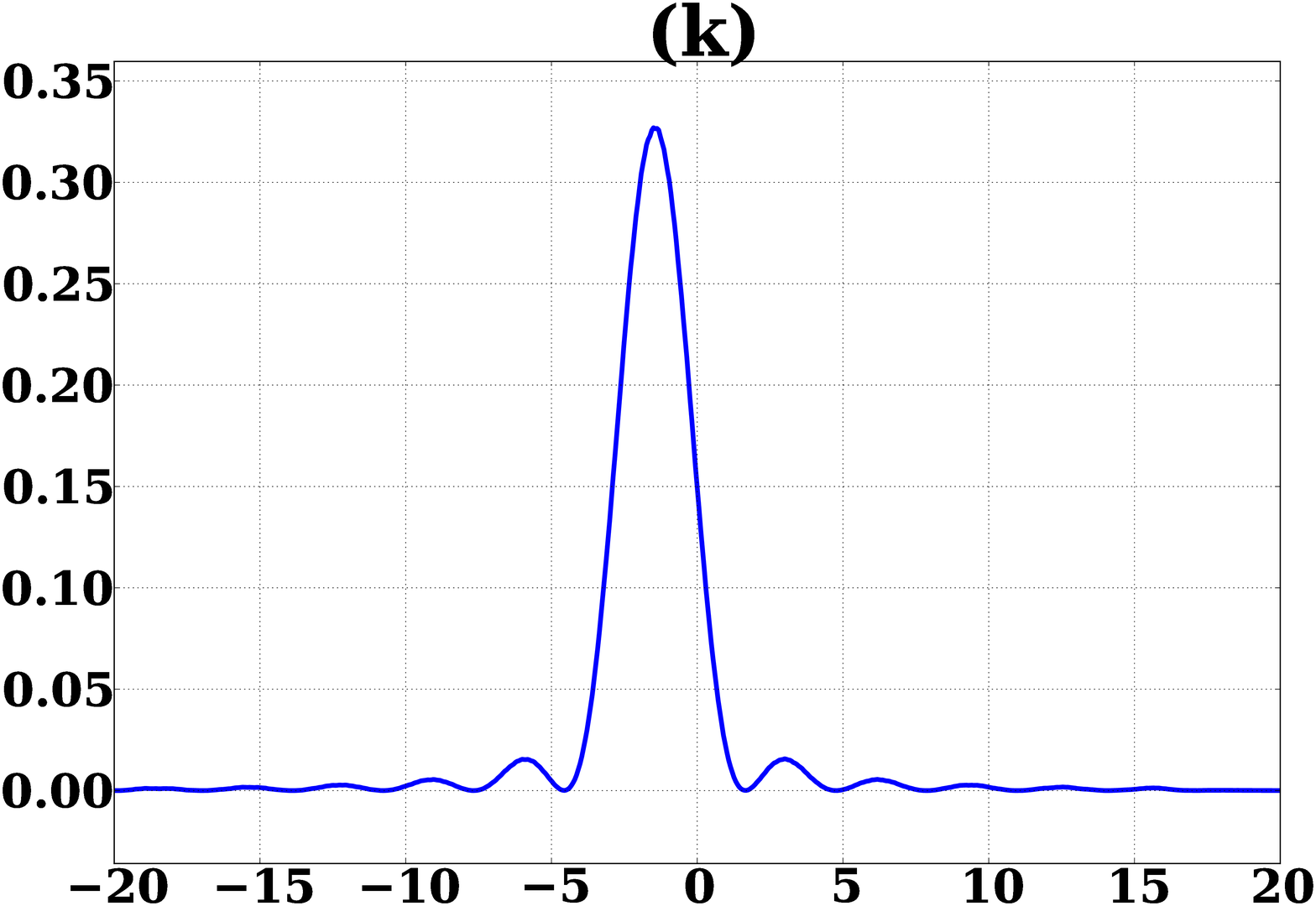}}
\resizebox {0.23 \textwidth} {0.172 \textheight }{  \includegraphics{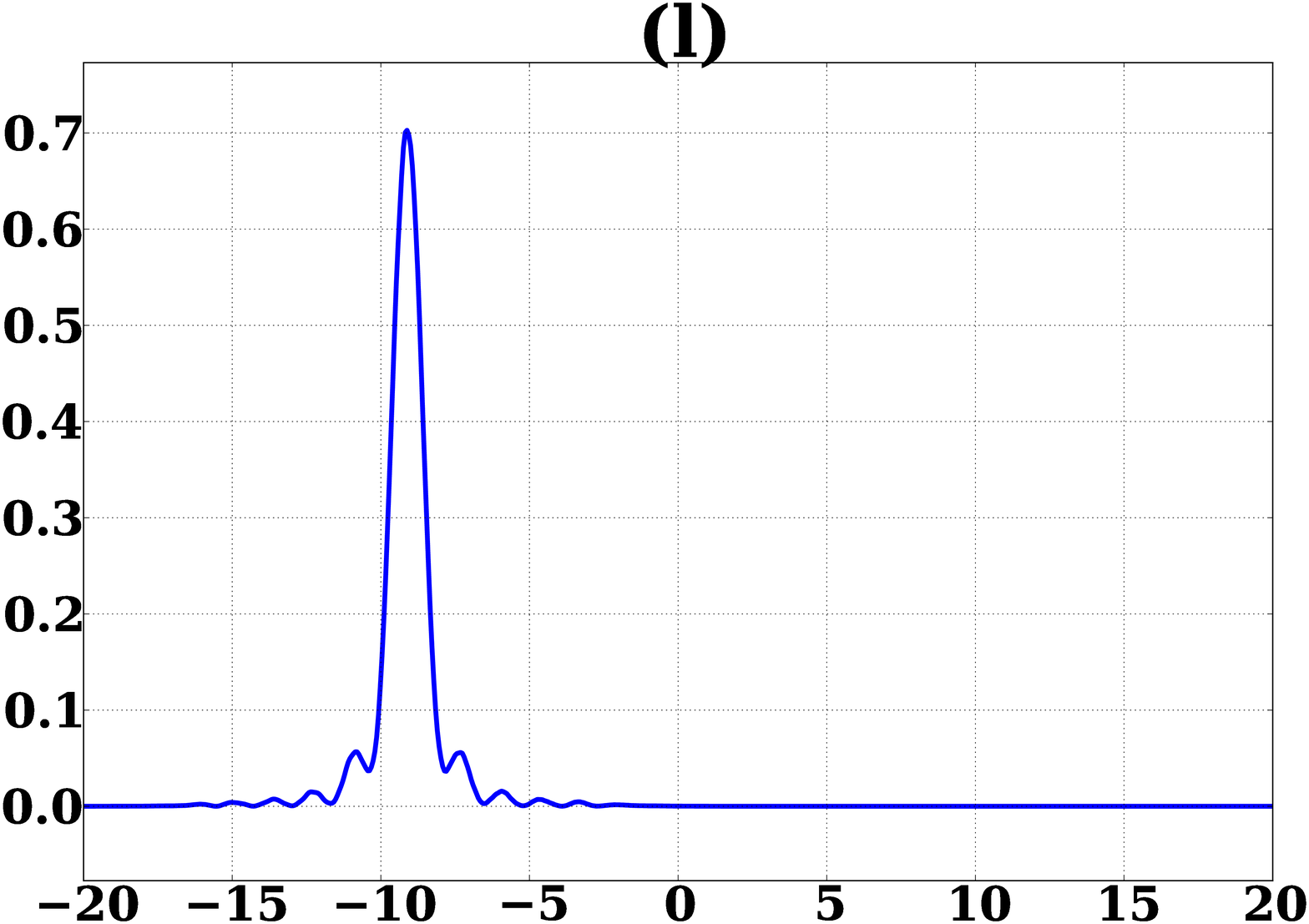}} 
\caption{Probability distribution for location states of a harmonic oscillator for $T$ = (a) 0.0, (b) 1.0, (c) 2.0, (d) 3.0, (e) $\pi$, (f) 4.0, (g) 5.0, (h) 6.0, (i) $2\pi$, (j) 7.0 (k) 8.0 (l) 9.0.}
\label{fig:pmic_shm_Oscill}         
\end{figure}

For the above values of  parameters, the period of oscillation of the harmonic oscillator is $2\pi$. We have also plotted the  patterns for a full period, at values of $t$ = 0.0, 1.0, 2.0, 3.0, $\pi$, 4.0, 5.0, 6.0, $2\pi$  7.0, 8.0 and 9.0  in Fig. \ref{fig:pmic_shm_Oscill}. It can be  clearly seen that the probability density of location state is periodic and maintains the diffraction pattern as it moves along, changing from the Fresnel to Fraunhofer  and vice-versa, in this case.   The oscillatory behaviour has some  similarity with that of coherent states, and it  indicates that  the characteristics of location states deserves to be studied in detail.

\section{Quantum treatment of diffraction} \label{sec:diffraction}

Now we turn to the actual quantum treatment of diffraction of free particles when they pass through a single slit of width $a$ made on a diaphragm placed in a region where there is no other potential. The experimental set-up is as described in Sec. \ref{sec:review}. Let the particle be incident on the diaphragm from left, and the incident wave function  be of the form $\Psi_x(x,t)\Psi_y(y,t)\Psi_z(z,t) = N e^{i(k_xx-E_x t/\hbar)}$. This plane wave is assumed to have a constant wave vector ${\vec{k}}$, whose components are   $k_x>0$, $k_y=0$, $k_z=0$, so that $\Psi_y$ and $\Psi_z$ are constants and the initial wave function corresponds to a wave  progressing along the positive $x$-axis. We assume the collapse to occur at $t=0$, when the particle passes through the slit. The part $\Psi_x(x,t)$ of the collapsed wave function is  not affected. For $t\geq 0$, the part $\Psi_y(y,t)$ is assumed to be given by equation (\ref{eq:slit_wavefn4}), which is the location state along the $y$-axis. The  product wave function that describes the particle  is now 

\begin{equation}
\Psi_x(x,t)\Psi_y(y,t)= \frac{N\; e^{i(k_xx-E_x t/\hbar)}}{2\pi \sqrt{a}}  \int_{-k_{m}}^{k_{m}}dk_y  \left[ \int_{-a/2}^{a/2}dy^{\prime} e^{-ik_y y^{\prime}} \right]   e^{i(k_y y-E_{y} t/\hbar)}, \label{eq:comb_wave_fn}
\end{equation}
where $E_{y}=\hbar^2k_y^2/2m$. To obtain  the diffraction pattern  on the screen placed at $x=D$, one has to plot the modulus square of this product function on  the screen. We do not consider a factor  $\Psi_z(z,t)$ in the above product since the free incoming particle has $k_z=0$ and also since the slit is  of infinite depth along the $z$-axis. Thus  $\Psi_z(z,t)$  will always remain a constant. 

It is easily seen that  $|\Psi_x(x,t)\Psi_y(y,t)|^2 $ does not depend on $x$. The only variation for this probability density is along the $y$-direction and for different values of $t$, the patterns change as  in Fig. \ref{fig:pmic_free}.  Hence the probability density on a screen evaluated using the above wave function shall be  independent of the position $x=D$  of the screen, but will depend on time $t$ for a fixed $D$.  On the other hand,  the patterns obtained  in  experiments depend on $D$, but for a fixed $D$, they are time-independent. Thus unless we specify some time $T$ in the above expression corresponding to a given value of $D$, there is no definite theoretical prediction of the pattern on the screen to compare with  experiment. It is clear that if we adhere to standard   quantum mechanics alone, some sort of  discrepancy arises in this case.

The  way out of this puzzle is to assume that there are point particles moving along trajectories, as in  nonlocal hidden variable theories. The  dBB, MdBB and the FFM formulations are such hidden variable theories.  In the above case of single slit diffraction, the particles move with a constant $x$-component of velocity $v_x=\hbar k_x/m$   in all the three formalisms. (Though the FFM case differ with the former two cases with regard to time parametrization \cite{floyd1,matone1},  the free particle motion in it is identical with that in the former ones.) Therefore the time  with which a particle from the slit  reaches the screen placed at $x=D$ is $T=D/v_x$.  We shall now attempt to plot the probability distribution versus $y$, on this screen at $D$ for a   time evaluated according to this formula.  The plots shown in Fig. \ref{fig:fresnel_fraun_patterns}    are  for a slit of width $a=0.1$.

\begin{figure}[h]
\centering 
\resizebox {0.45 \textwidth} {0.3 \textheight } {\includegraphics{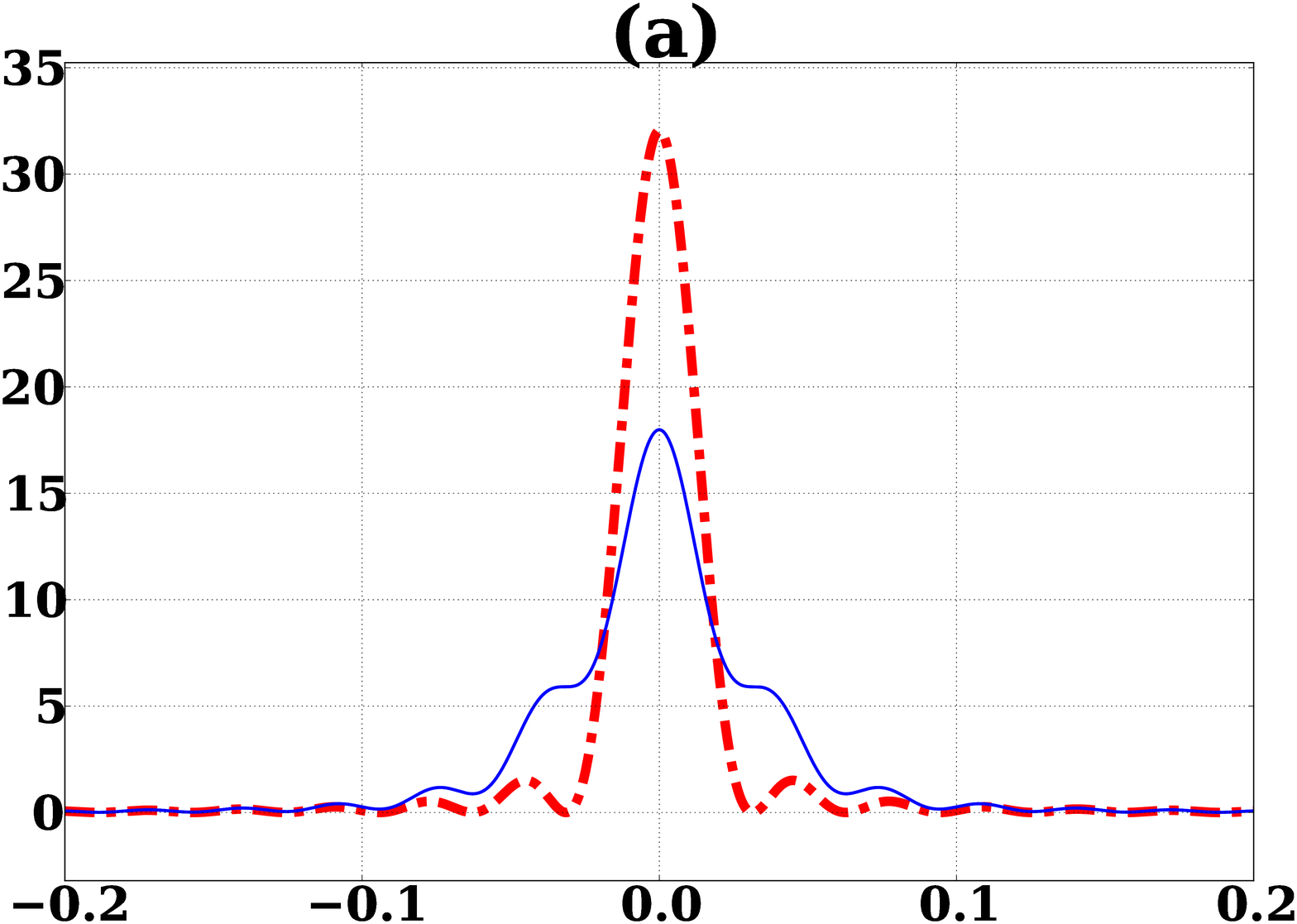}} 
\resizebox {0.45 \textwidth} {0.3 \textheight } {\includegraphics{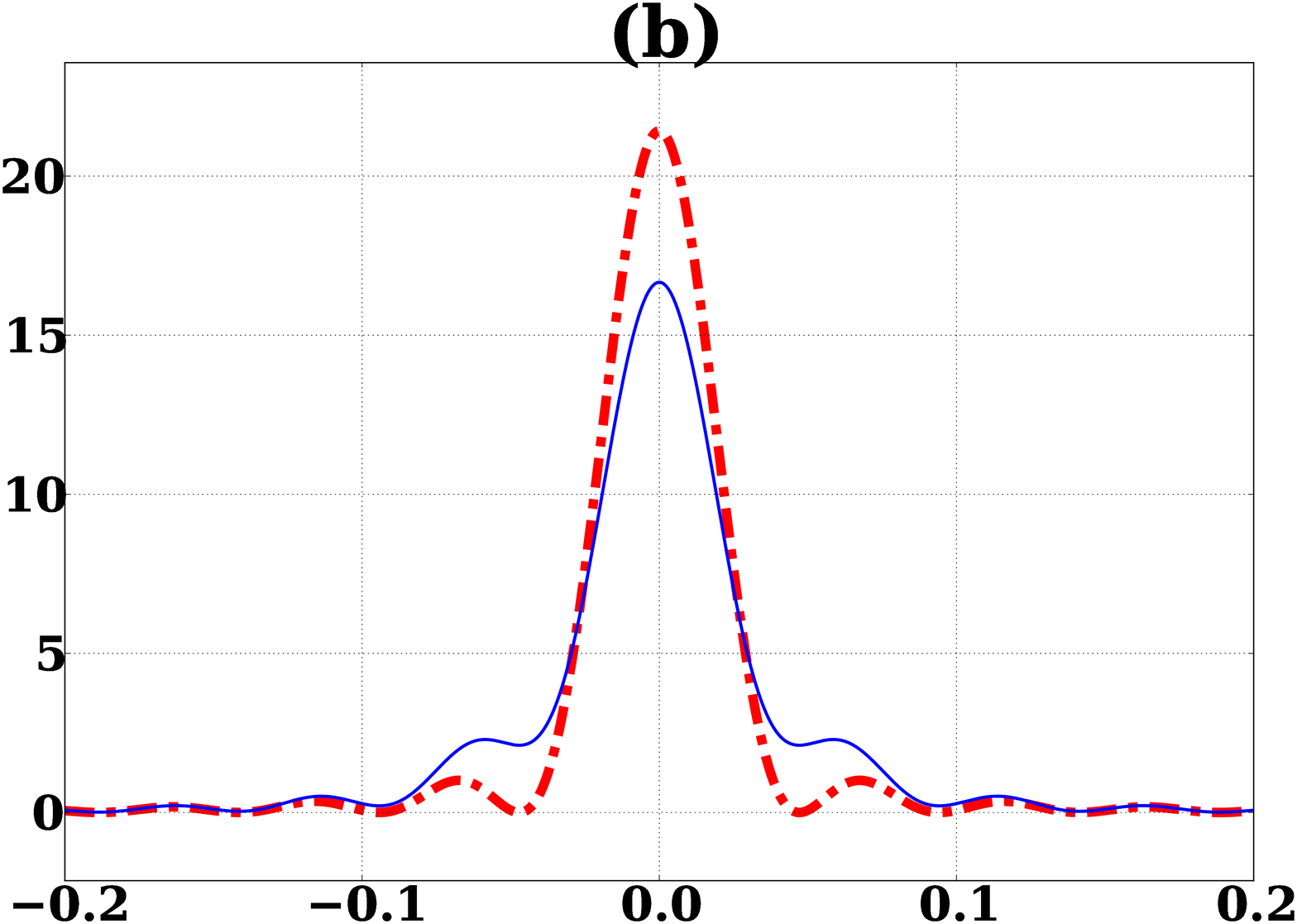}}
 \resizebox {0.45 \textwidth} {0.3 \textheight } {\includegraphics{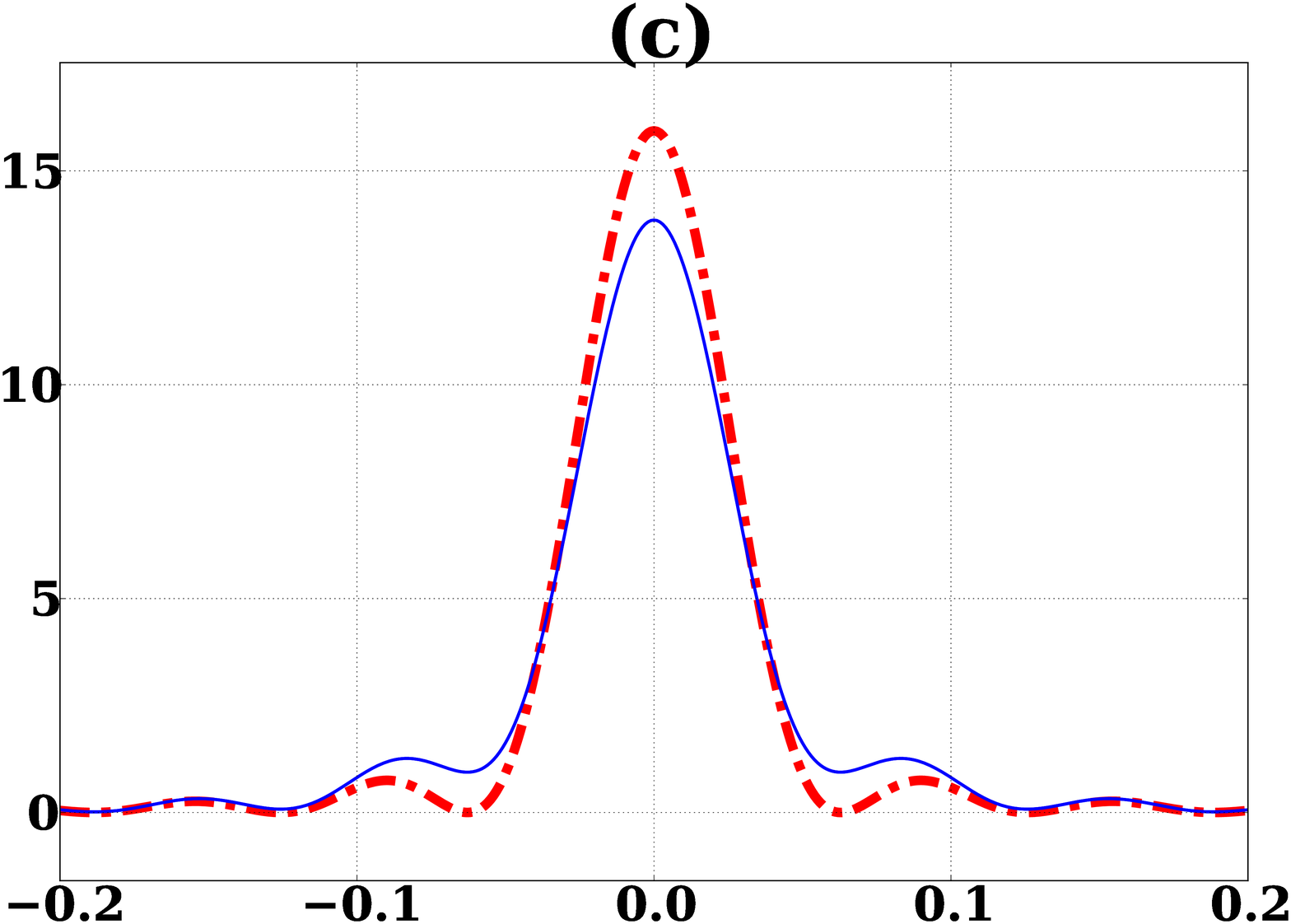}} 
  \resizebox {0.45 \textwidth} {0.3 \textheight} {\includegraphics{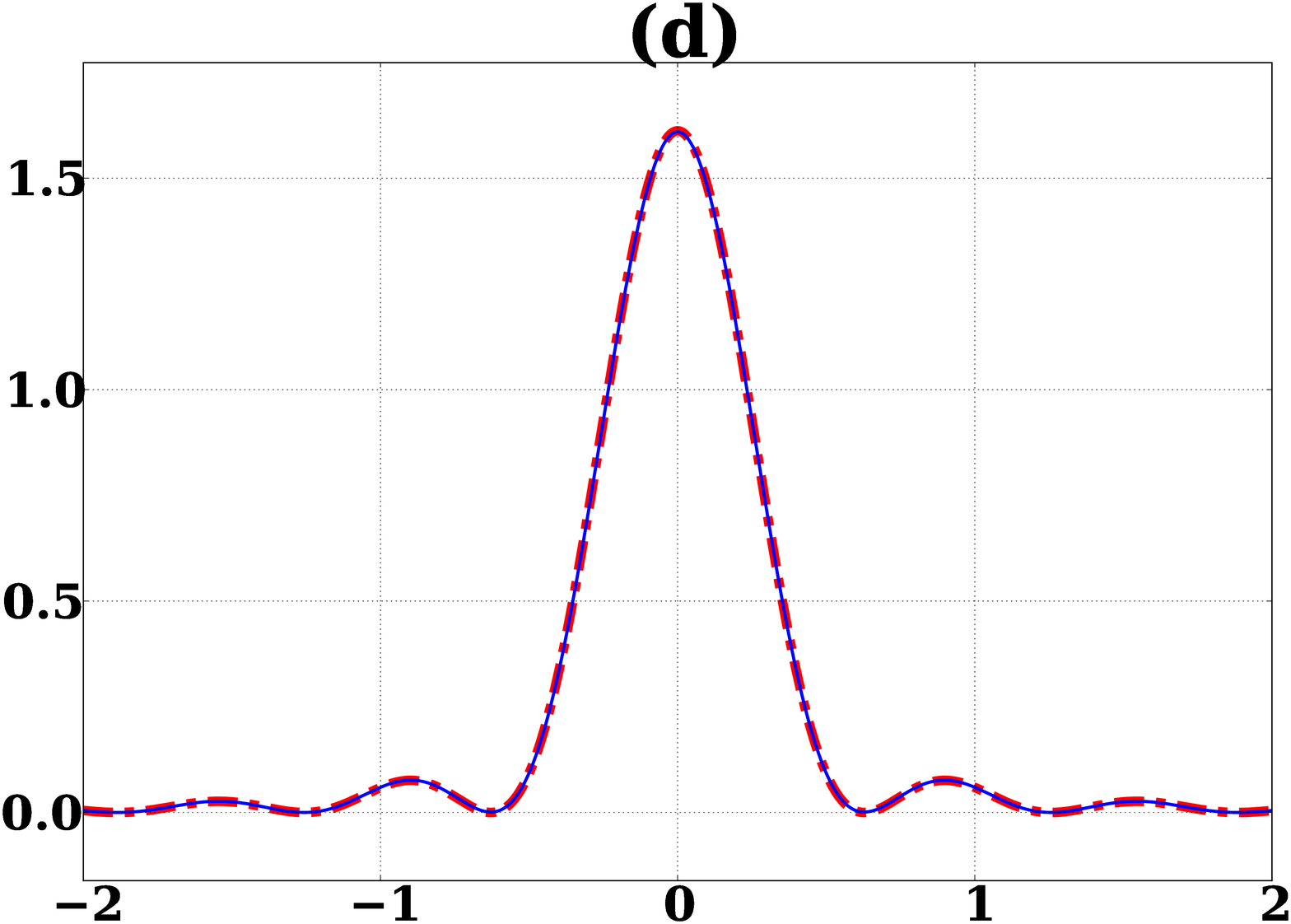}}
\caption{Diffraction patterns predicted according to the location state formalism (blue lines) when compared to that in \cite{marcella} (dotted red line) for values of $T$= (a) 0.0005,   (b) 0.00075, (c) 0.001, (d) 0.01}
\label{fig:fresnel_fraun_patterns}        
\end{figure}

In all the figures in Fig. \ref{fig:fresnel_fraun_patterns}, we have plotted  $|\Psi_x(x,t)\Psi_y(y,t)|^2 $, which is the probability distribution  corresponding to Eq. (\ref{eq:comb_wave_fn}) (blue lines), together with the     expression (\ref{eq:std_fraun}) used by Marcella \cite{marcella} (dotted red lines), both normalised and evaluated for fixed values of $D$.  The different values of $D$, along with that of $a$ and $\lambda$, give  Fresnel numbers $N_F\equiv \frac{a^2}{4\lambda D}=\frac{a^2m}{8\pi\hbar T}$ =  0.796, 0.531, 0.398, 0.039, respectively. This shows that except for the last case, the diffraction is  in the Fresnel region. Our plots using Eq. (\ref{eq:comb_wave_fn}) demonstrate that at  small values of time $T$ (closer distances $D$), the location state pattern approaches that of Fresnel diffraction, contrary to what happens in \cite{marcella}.  For the last plot, we have $N_F<<1$ and hence the Fraunhofer regime results.  The agreement in the last case can be seen to be excellent. Thus the comparison between  patterns of the present location state formalism and that in \cite{marcella} tells that the present one has better agreement with experiment in all regions.

We thus see that the standard results of Fresnel and Fraunhofer diffraction can be connected to the time-evolution of location states only with the help of quantum trajectory formalisms. Our success in this endeavour     supports  the existence of particle trajectories. In this case,  dBB, MdBB and FFM trajectories give the same results, for they  have the same value for $v_x=\hbar k_x/m$. Therefore, it is not possible to discriminate them using this experiment.

\section{Discussion} \label{sec:discussion}

 The present attempt to describe quantum mechanically the phenomenon of diffraction is strongly founded on the collapse of a quantum state due to  position measurement and the subsequent time-evolution of this   state. In dealing with this time-evolution,  we have  followed the standard axioms of quantum mechanics. That these patterns (for instance, those in section \ref{sec:free})    closely resemble the diffraction patterns obtained in experiments is reassuring and shows the connection of  diffraction to the quantum location states, which we have defined in this paper. Since it is here shown possible that a single quantum mechanical  expression can explain both the Fresnel and Fraunhofer diffractions, the present formulation deserves  serious attention.

To obtain the diffraction pattern  using the above formalism, we must take the limit $k_m \rightarrow \infty$ or $n\rightarrow \infty$, in the respective integrations or summations done to evaluate $\Psi_y(y,t)$ in equations (\ref{eq:slit_wavefn4}) and (\ref{eq:slit_wavefn6}). We have drawn the patterns  with finite values of these parameters, but  it is observed that they  do not get modified appreciably by increasing these values beyond that  used by us.   However, one must realise that in such limits, the mean value of energy in these states also must tend to infinity. It is only natural in the quantum regime that a position eigenstate, which is described by a Dirac $\delta$-function, involves infinite energy, for the uncertainty in momentum tends to infinity in those cases.  The location states are written as an integral over such position eigenstates, with eigenvalues lying between $-a/2$ to $a/2$. We have evaluated the mean value of energy of these location states for various  $k_m$ only to find that the mean  keeps on increasing. Since  the variance of momentum in the state (\ref{eq:slit_wavefn1}) is infinite, an infinite mean value of energy is unavoidable. This peculiarity is a consequence of  the discontinuity of this wave function and is characteristic of all orthogonal measurements \cite{kip}. But in actual experiments, such sharp discontinuities may not exist, so that one can have a finite upper limit for $k_m$ or $n_m$, as we have taken. In any case, since the  patterns do not change appreciably with their values beyond those used by us, our explanation of diffraction  remains satisfactory. However,  more sophisticated experiments are needed to settle the issue of energy.

 It is true that both  formalisms, i.e., the one  in Ref. \cite{marcella} and the present  one, use some kind of hidden variable approaches. In the former case, while using $p_y =p\sin \theta$, it is assumed that the particles are emitted with definite values of $p_x$ and $p_y$, and that they proceed to the screen as classical particles with precisely these momenta. In fact, no time-evolution of the wave packet for $t>0$ is  considered here. Hence \cite{marcella} cannot be considered as relying on a proper hidden variable theory.  One must also note that  the formalisms in \cite{marcella} and \cite{fabbro} do not focus on Fresnel diffraction. The present one, on the other hand, considers the evolution of wave function and makes use of standard nonlocal hidden variable theories such as those in dBB, MdBB and FFM formalisms. Taking into account the evolution of the state makes the formalism capable of providing a satisfactory explanation even to the  Fresnel diffraction,  at small distances from the slit.  While showing that the location states play a major role in providing this unified description of  Fraunhofer and Fresnel diffractions, we  find this result also as suggestive of the existence of quantum trajectories. 
 
\section*{Acknowledgements}
We thank Professors E.R. Floyd, K. Babu Joseph and M. Matone for insightful discussions and Arunkumar A. for his valuable help in computations.




\begin{thebibliography}{99}

\bibitem{epstein}  Epstein, P.S.,  Ehrenfest,  P.: The quantum theory of the Fraunhofer diffraction. Proc. Nat. Acad. Sci. \textbf{10}, 133 (1924)

\bibitem{sudarshan} Sudarshan, E.C.G.,  Rothman, T.: The two-slit interferometer reexamined.  Am. J. Phys. \textbf{59}, 592 (1991)

\bibitem{marcella} Marcella, T.V.: Quantum interference with slits. Eur. J. Phys. \textbf{23}, 615 (2002)

\bibitem{rothman} Rothman, T., Boughn, S.: `Quantum interference with slits' revisited. Eur. J. Phys. \textbf{32}, 107 (2011)

\bibitem{fabbro} Fabbro, B.: On the quantum theory of diffraction by an aperture and the Fraunhofer diffraction at large angles. arXiv:1710.09758v2 [quant-ph] (2018)

\bibitem{kip} Braginski, V.B., Khalili, F.Y.: Quantum measurement. Ed. Kip S. Thorne. \newblock Cambridge University Press  (1992)

\bibitem{wise} Wiseman,  H.M.,  Milburn, G.J.: Quantum Measurement and Control. \newblock Cambridge University Press (2010)

\bibitem{bell} Bell, J.S.: Speakable and unspeakable in quantum mechanics. \newblock Cambridge University Press (1987)



\bibitem{dBB1}
de-Broglie, L.:  La m{\'e}canique ondulatoire et la structure atomique de la
  mati{\`e}re et du rayonnement (Wave mechanics and the atomic structure of
  matter and radiation).  J. Phys. Radium, \textbf{8}, 225 (1927)
  
  \bibitem{dBB2}
Bohm, D.: A suggested interpretation of the quantum theory in terms of
  hidden  variables. 1.  Phys. Rev. \textbf{85}, 166 (1952)



\bibitem{MdBB}
John, M.V.: Modified de Broglie-Bohm approach to quantum mechanics.
  Found. Phys. Lett. (1988 - 2006), \textbf{15}, 329 (2002)

\bibitem{yang}
Yang, C.-D.: Quantum dynamics of hydrogen atom in complex space. Ann. Phys.\textbf{319}, 399 (2005)  

\bibitem{tannor} Goldfarb, Y., Degani, I., Tannor, D.J.: Bohmian mechanics with complex action: a new trajectory-based formulation of quantum mechanics. J. Chem. Phys. \textbf{125}, 231103 (2006)



  \bibitem{floyd}
Floyd, E.R.: Modified potential and Bohm's quantum-mechanical potential.
  Phys. Rev. D, \textbf{26}, 1339 (1982)


  
  \bibitem{matone} Faraggi, A.E., Matone, M.: The equivalence postulate of quantum mechanics. Int. J. Mod. Phys. A, \textbf{15} 1869 (2000)
  



\bibitem{sakurai} Sakurai, J.J.: Modern quantum mechanics. \newblock Pearson Education Inc. (1994)

\bibitem{gottfried} Gottfried, K.:  Quantum Mechanics, Vol. I: Fundamentals. \newblock Benjamin (1966) 

  \bibitem{floyd1} Floyd, E.R.: Action quantization, energy quantization, and time parametrization. Found. Phys. \textbf{47}, 392 (2017)
  
  \bibitem{matone1}  Faraggi, A.E., Matone, M.: Hamilton-Jacobi Meet Mobius, J. Phys, Conference Series,  \textbf{ 631},  12010 (2015)

\end{thebibliography}
\end{document}